\documentclass[acmsmall,authorversion]{acmart}

\usepackage{cleveref}
\usepackage{bussproofs}
\usepackage{tikz}
\usepackage{stmaryrd}
\usepackage{microtype}

\setcitestyle{nosort}
\newcommand{\newcommenter}[3]{\newcommand{#1}[1]{}}

\newcommenter{\lk}{blue}{LK}
\newcommenter{\jmc}{red}{JMC}
\newcommenter{\plr}{magenta}{PLR}

\newcommand{\happensbefore}[0]{\textsf{happens-before}}

\setcopyright{rightsretained}
\acmDOI{10.1145/3649830}
\acmYear{2024}
\copyrightyear{2024}
\acmSubmissionID{oopslaa24main-p60-p}
\acmJournal{PACMPL}
\acmVolume{8}
\acmNumber{OOPSLA1}
\acmArticle{113}
\acmMonth{4}
\received{2023-10-20}
\received[accepted]{2024-02-24}

\begin{document}

\title{Inductive Diagrams for Causal Reasoning}

\author{Jonathan Castello}
\orcid{0000-0002-8548-3683}
\affiliation{%
  \institution{University of California, Santa Cruz}
  \country{USA}}
\email{jcaste14@ucsc.edu}

\author{Patrick Redmond}
\orcid{0000-0001-5702-0860}
\affiliation{%
  \institution{University of California, Santa Cruz}
  \country{USA}}
\email{plredmond@ucsc.edu}

\author{Lindsey Kuper}
\orcid{0000-0002-1374-7715}
\affiliation{%
  \institution{University of California, Santa Cruz}
  \country{USA}}
\email{lkuper@ucsc.edu}

\begin{abstract}
  The Lamport diagram is a pervasive and intuitive tool for informal reasoning about ``happens-before'' relationships in a concurrent system.
  However, traditional axiomatic formalizations of Lamport diagrams can be painful to work with in a mechanized setting like Agda. 
  We propose an alternative, inductive formalization --- the \emph{causal separation diagram} (CSD) --- that takes inspiration from string diagrams and concurrent separation logic, but enjoys a graphical syntax similar to Lamport diagrams.
  Critically,  CSDs are based on the idea that causal relationships between events are witnessed by the \emph{paths} that information follows between them.
  To that end, we model ``happens-before'' as a dependent type of paths between events.

  The inductive formulation of CSDs enables their \emph{interpretation} into a variety of semantic domains.
  We demonstrate the interpretability of CSDs with a case study on properties of \emph{logical clocks}, widely-used mechanisms for reifying causal relationships as data.
  We carry out this study by implementing a series of interpreters for CSDs, culminating in a generic proof of Lamport's \emph{clock condition} that is parametric in a choice of clock.
  We instantiate this proof on \citeauthor{lamport-clocks}'s scalar clock, on \citeauthor{mattern-vector-time}'s vector clock, and on the matrix clocks of \citeauthor{rst-matrix-clock} and of \citeauthor{wuu-bernstein-matrix-clocks}, yielding verified implementations of each.
  The CSD formalism and our case study are mechanized in the Agda proof assistant.
\end{abstract}



\maketitle

\section{Introduction}
\label{sec:intro}

Causality --- the principle that an effect cannot precede its cause --- is of central importance in concurrent and distributed systems. It undergirds every protocol for strengthening message-passing communication models beyond asynchrony, and it allows the concept of a sequence of actions or flow of messages to be well-defined in the first place.
Verification of properties related to causality, such as the causal consistency of data stores~\citep{lesani-chapar,gondelman-distributed-causal-memory} or the causal order of message delivery~\cite{nieto-verified-crdts,redmond-verified-causal-broadcast}, inevitably requires the modeling of some notion of ``history'' --- for instance, a per-process log of received messages in receipt order --- against which causally-sensitive properties can be judged.
These representations of history originate in the process model of \citet{lamport-clocks} and its associated \emph{\happensbefore{}} relation, a concrete representation of causal relationships amongst events in a system.
In this model, a history is given by a sequence of primitive events for every participating process, together with a visibility relation pairing events across processes (as with pairs of \emph{send} and \emph{receive} events).
The \happensbefore{} relation is derived from this data, as the transitive closure of the process sequences together with the visibility relation.

Lamport-style models of executions are intuitive, elegant, and ubiquitous.
However, they are typically characterized purely \emph{axiomatically} rather than \emph{inductively}.
While this makes them well-suited to traditional pencil-and-paper proofs, our experience has been that applying them to \emph{mechanized} proof is a considerable struggle.
Axiomatic, set-theoretic models do not always translate cleanly into constructive type theory, and the resulting encodings may not take the best advantage of the tools at hand.
The resulting representations of history lead to tedious and ad-hoc proofs, due to the need to reason about causality as a derived notion rather than a fundamental one.
However, by baking causal information more deeply into the data model of histories, we can obtain a causally-directed induction principle that eliminates much of the tedium.

To that end, in this paper we develop a novel \emph{inductive} representation of history --- tailored to the needs of verification in a mechanized setting --- in which causal information is immediately at hand.
We propose \emph{causal separation diagrams} (CSDs), which are intended to serve as a drop-in replacement for existing Lamport-style models of history, such as when considering a program or protocol in terms of its possible execution histories.
Where Lamport-style histories must be proved acyclic to be physically meaningful, CSDs are automatically acyclic by construction.

\paragraph{Verification by interpretation.}
A CSD represents a particular execution of a program (or protocol).
As a means of representing executions, CSDs are not tied to any particular verification methodology.
To verify properties of the program as a whole, we can reason in aggregate about the executions that can arise from running it, in the same way Lamport-style executions are commonly used.
However, since CSDs are presented as an inductively-defined dependent data type, it is natural to give compositional \emph{interpretations} of CSDs into other data types.
This interpretability suggests a particular approach to verification in which the building blocks of executions are \emph{interpreted as proof steps}, then composed along their causal structure.

As a demonstration of the interpretability of CSDs, we consider the verification of \textbf{logical clocks}, a common class of devices for reifying causal information into a system at runtime~\citep{raynal-singhal-survey}.
In particular, the \textbf{clock condition}~\citep{lamport-clocks} is an essential property of logical clocks, assuring that two causally-ordered events are assigned like-ordered timestamps.
We mechanically verify the clock condition for a broad class of logical clocks, including Lamport clocks~\citep{lamport-clocks}, vector clocks~\citep{mattern-vector-time,fidge-vector-time,schmuck-dissertation}, and matrix clocks~\citep{wuu-bernstein-matrix-clocks,rst-matrix-clock}, by giving a series of interpreters for CSDs.
This interpretation-based approach to verification would be awkward and difficult without an inductive data structure that accounts for causality; but with CSDs, it becomes natural and straightforward.

In summary, the main contributions of this paper are as follows:

\begin{itemize}
\item \textbf{Causal separation diagrams (CSDs).} After presenting informal intuitions in \Cref{sec:csds-informally}, we describe a new formal diagrammatic language for reasoning about executions of concurrent systems (\Cref{sec:csds-defined}).  CSDs are inspired by Lamport diagrams --- a well-established visual language for expressing the behavior of distributed systems --- but they are inductively defined, which makes them amenable to interpretation into many semantic domains.

\item \textbf{Interpreting CSDs.} We present interpretations of CSDs into three semantic domains:
  \begin{itemize}
  \item \textbf{Into types:} We define an interpretation of CSDs into the domain of \emph{causal paths} (\Cref{sec:interp-causal-paths}).  Causal paths are a proof-relevant analogue of Lamport's \happensbefore{} relation, where any given path inductively describes a particular flow of information.

  \item \textbf{Into functions:} We define an interpretation of CSDs into a domain of \emph{clocks}; that is, functions that compute a logical timestamp at every event (\Cref{sec:interp-clock-programs}). Our interpretation is parametric in the particular choice of logical clock, so long as it is realizable as an abstract data type with \textbf{increment} and \textbf{merge} operations~(\Cref{subsec:realizable-clocks}).

  \item \textbf{Into proofs relating types and functions:} We relate the above interpretations via a third interpretation of CSDs into proofs that clocks respect causality (\Cref{sec:interp-proofs}). This yields a proof of Lamport's clock condition for any realizable clock whose timestamps increase with successive operations.
  \end{itemize}

\item \textbf{Applying CSDs: verified logical clocks.} Finally, we instantiate our interpretations on the clocks of \citeauthor{lamport-clocks}, \citeauthor{mattern-vector-time}, \citeauthor{rst-matrix-clock}, and \citeauthor{wuu-bernstein-matrix-clocks}, yielding mechanically verified implementations of each (\Cref{sec:verified-clocks}).  In particular, we give the first (to our knowledge) mechanized proofs of the clock condition for both matrix clocks.
\end{itemize}

All of our contributions are mechanized in the Agda proof assistant, and have been included in our open-source library for working with CSDs, available at \url{https://github.com/lsd-ucsc/csds}.

\section{From Informal Diagrams to Formal Models}
\label{sec:csds-informally}

\begin{figure}
  \includegraphics[width=\columnwidth]{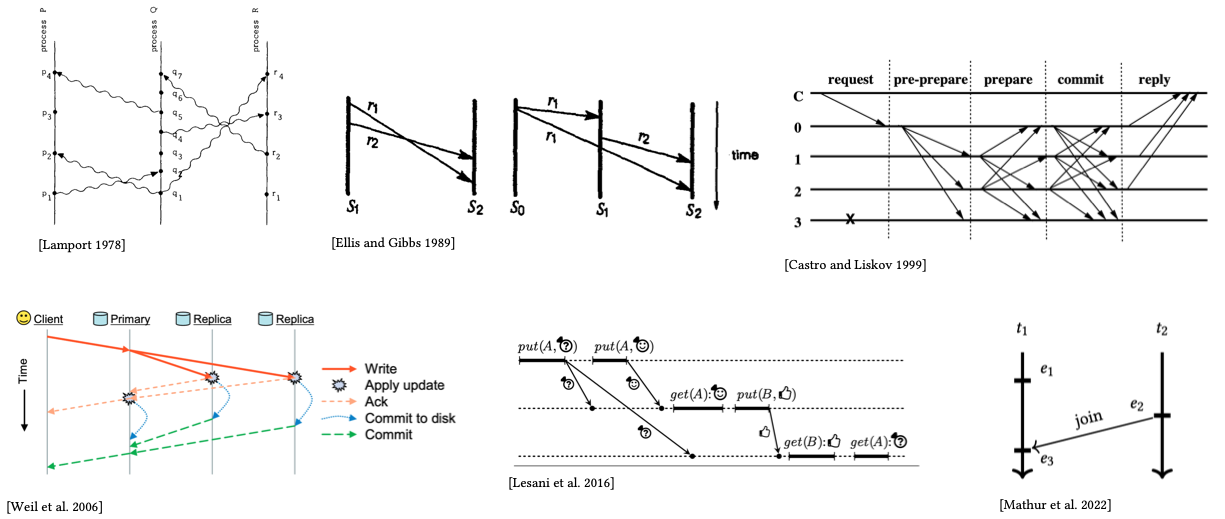}
  \caption{An assortment of Lamport diagrams from the literature.  In these examples, time flows from top to bottom~\citep{ellis-gibbs-ot,weil-ceph,mathur-tree-clocks}, from left to right~\citep{castro-liskov-pbft,lesani-chapar}, or, rarely, from bottom to top~\citep{lamport-clocks}, and parallel through-lines represent processes, threads, or spatially-separated sites, while arrows represent communication between them.}
  \label{fig:lamport-diagram-examples}
\end{figure}
Lamport diagrams
\footnote{Lamport diagrams go by many names, including time diagrams, spacetime diagrams, sequence diagrams, and more.  While \citet{lamport-clocks}'s analysis of causality in the context of distributed systems was an early use of such diagrams, it appears to not have been \emph{the} first in the published literature; the oldest we have found is via \citet{le-lann-distsys}.}
are a ubiquitous device for visualizing causal relationships over space and time; see \Cref{fig:lamport-diagram-examples} for a diverse selection spanning six decades of computing literature.
In a Lamport diagram, logically-separate processes evolve over time along straight through-lines: their actions are represented as dots (or similar) on a given process line, and their communications yield arrows crossing laterally between process lines.
Importantly, causal relationships are reduced to simple geometric paths: two points in space and time are causally ordered if, and only if, they are connected by a forward path along the diagram.

As illustrations, Lamport diagrams are by nature informal.
To support \emph{formal} reasoning about concurrent systems, we need formal models that capture the same scenarios displayed by these diagrams.
\citet{lamport-clocks} presented the following model, a natural generalization of sequential processes to multiple participants.
\begin{definition}[Lamport execution \citep{lamport-clocks}]
  \label{def:lamport-execution}
  A \textit{Lamport execution} is:
  \begin{itemize}
    \item A set \(P\) of \emph{processes}, each of which is a sequence of atoms called \emph{actions}\footnote{We avoid the traditional term ``event'', for now, because the causal relation we define in \Cref{sec:interp-causal-paths} only indirectly relates actions. A causal order ought to be defined on ``events''; so we reserve that term and speak of ``actions'' here instead.}; together with

    \item A set \(M\) of \emph{messages}, each of which is an ordered pair of actions across two processes (the message's associated ``send'' and ``receive'' actions).
  \end{itemize}
\end{definition}
\begin{definition}[Happens-before \citep{lamport-clocks}]
  \label{def:happens-before}
  Given a Lamport execution, the \emph{\happensbefore{}} relation on actions, written \(a_1 < a_2\), is the transitive closure of the execution's set of messages together with the total orders given by each process.

  By tradition, executions for which \happensbefore{} is not a (strict) partial order (i.e. fails to be asymmetric) are excluded from consideration, as these indicate a failure of causality.
\end{definition}

\begin{figure}
  \begin{center}
    \includegraphics[width=0.45\columnwidth]{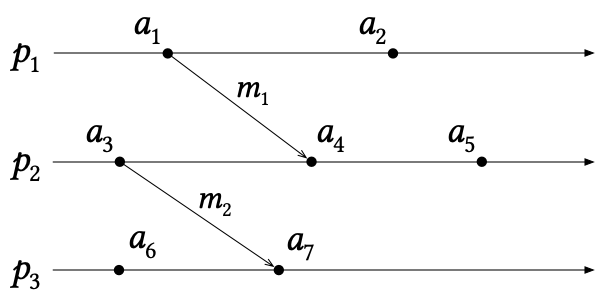}%
  \end{center}
  \caption{An example Lamport diagram.}
  \label{fig:lamport-diagram-example}
\end{figure}
The data of a Lamport execution can be visualized by a Lamport diagram.
For example, the Lamport diagram in \Cref{fig:lamport-diagram-example} depicts an execution involving three processes, \(p_1\), \(p_2\), and \(p_3\), each having performed a few actions.
Some of the actions in this execution are causally ordered: we see that \(a_1 < a_4\) since \(a_1\) and \(a_4\) are the send and receive actions of message \(m_1\), and \(a_4 < a_5\) because they occur in sequence on \(p_2\).
Therefore, by transitivity, \(a_1 < a_5\).
We also have that \(a_3 < a_4\) and \(a_3 < a_7\), among other relationships.
However, \(a_1\) and \(a_3\) are not related by \happensbefore{}, nor are \(a_4\) and \(a_7\); such pairs are said to be \emph{concurrent} (or \emph{causally independent}).

\begin{figure}
  \begin{center}%
    (a)
    \includegraphics[width=0.40\columnwidth]{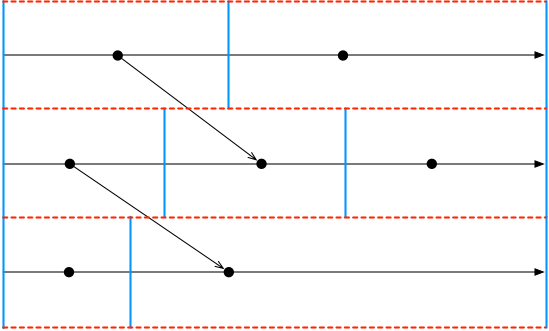}%
    \qquad%
    (b)
    \includegraphics[width=0.40\columnwidth]{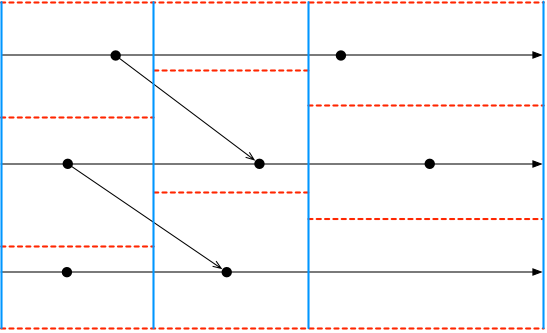}%
  \end{center}
  \caption{Two ways to decompose the Lamport diagram of \Cref{fig:lamport-diagram-example} into ``tiles''.  On the left (a), we split first along spatial boundaries (dashed red lines), yielding individual processes, and then along temporal boundaries (solid blue lines). On the right (b), we split first along temporal boundaries, yielding consistent cuts, and then along spatial boundaries.}
  \label{fig:diagram-decompositions}
\end{figure}

It is also possible to go the other way, taking a Lamport diagram and formalizing the scenario it displays as a Lamport execution.
Therefore, we can consider the diagram to come first, with the derivation of a formal execution from an informal diagram serving as an origin story for the formal model itself.
We can rederive the traditional execution by first splitting a diagram along spatial boundaries --- separating the process lines from one another --- and then separating the sequential actions along each process line by temporal boundaries.
Doing so for the diagram in \Cref{fig:lamport-diagram-example} yields the decomposition in \Cref{fig:diagram-decompositions}(a).
However, we could also have begun by laying down a sequence of \textit{temporal} boundaries -- demarcating \emph{global steps} over the entire system -- and only then separating the atomic steps within each global step by spatial boundaries.
This approach might yield the decomposition in \Cref{fig:diagram-decompositions}(b).

Both decompositions yield a partition of the diagram into graphical tiles; and it is precisely the relationships between these tiles, witnessed by the dataflow lines passing between them, which must be captured formally.
In the traditional decomposition in \Cref{fig:diagram-decompositions}(a), tiles may be related across both temporal and spatial boundaries.
Process orders record the relationships across temporal boundaries, while messages record relationships across spatial boundaries.
\footnote{Depending on the execution being visualized, we may need to draw message-lines passing through tiles which neither send nor receive them; an effective visualization would be decidedly non-planar. Nonetheless, we consider that the relationship remains one of passing through the spatial medium.}
This data, comprising a traditional formal execution, is sufficient to capture all information presented in the diagram.

The state of affairs for our alternative decomposition in \Cref{fig:diagram-decompositions}(b) is notably different.
First, information flows between tiles only at temporal boundaries; spatial boundaries only separate causally-independent actions which cannot influence each other.
Intuitively, it takes time to move through space -- spatial boundaries separate actions which may as well occur simultaneously, so the propagation of information from one place to another can only occur across temporal boundaries.
However, this also means that differing quantities of state can leave a global step than enter it: a process may consume a message to decrease the quantity of data floating around, or emit a message to increase the quantity of data.
Without bracing ourselves against the suggestive global geometry of fixed parallel lines for each process, we cannot even distinguish process state from message state: a global step simply transforms one configuration of separated state into another.
Because of this indistinguishability, instead of referring to ``processes'' and ``messages'' we will refer only to \textbf{sites}: a site is a \emph{place where state exists}, encompassing both processes and messages.

Second, we could have drawn different temporal boundaries --- different \emph{consistent cuts} --- and found a different decomposition.
Consistent cuts~\citep{mattern-vector-time,chandy-lamport-snapshots} are of fundamental importance to the analysis of concurrent systems, as they model the realizable \emph{global states} of a system.
Thus, the formal representation for a diagram will embed a choice of consistent cuts; and as we will find in \Cref{sec:interp-clock-programs,sec:interp-proofs}, working with global information from the start enables simpler proof methods for reasoning about concurrent systems.
\footnote{We expect there to be a means of algebraically transforming a CSD to manipulate which consistent cuts it embeds; this would then yield a completely syntactic account of consistent cuts. However, we defer this to future work.}

Process lines can be recovered as chosen paths spanning the diagram --- that is, a chosen total order of actions, just as in the traditional execution.
These path essentially names pieces of state as they evolves over time; any state not on some path is, morally, a message.
We can even interpret this in a shared-memory setting: the configuration of sites along a consistent cut describe a shared heap, with each individual site modeling an exclusive region of memory.
A global step then updates the heap, claiming regions by merging them and releasing regions by splitting them apart.

\begin{figure}
  \begin{center}
    \includegraphics[width=0.40\columnwidth]{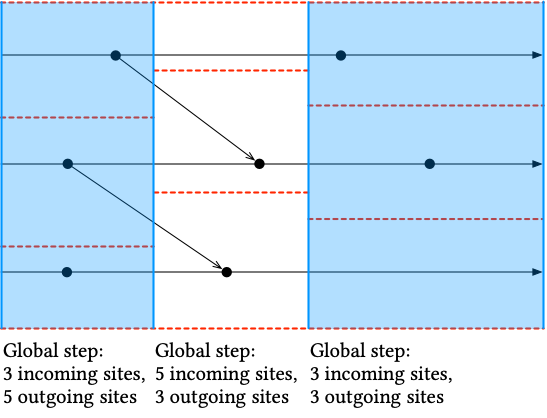}%
  \end{center}
  \caption{Global steps in our example diagram, with a site implied everywhere a cut intersects the diagram.}
  \label{fig:csd-decomposition-detail}
\end{figure}

\Cref{fig:csd-decomposition-detail} illustrates this notion of sites in more detail for our example.
The shaded global step on the left has three incoming sites and five outgoing sites, so we might compactly say it has type \(3 \rightrightarrows 5\) (``three to five''). The next two global steps have types \(5 \rightrightarrows 3\) and \(3 \rightrightarrows 3\), respectively.
Adjacent global steps must ``match up'' the sites on their incident site configurations; but during a global step, sites may be joined with or forked from others.

In \Cref{sec:csds-defined}, we will describe a novel formal model for concurrent executions based on these observations.
However, we can already see the shape this formalization must take:
\begin{itemize}
  \item
    Since we have essentially \textit{transposed} the sequential and concurrent boundaries compared to the traditional formalization, our formal data will consist of a sequence of global steps acting over separated state.

  \item
    Each global step will decompose into a collection of concurrent, atomic steps, no two of which act over the same site --- data flowing into and out of a global step must flow through precisely one of its constituent atomic steps.
    These steps include individual local actions \(a_1\), but also include fork actions (which split one site into two) and join actions (which fuse two sites into one).

  \item
    A causal relationship between actions \(a_1 \rightsquigarrow a_2\) will be witnessed by a sequence (or \emph{path}) of atomic steps, running forward from \(a_1\) to \(a_2\), such that adjacent steps share a site.
\end{itemize}

Our unification of messages and processes into sites makes our formalization ``natively'' suited for reasoning about shared-memory concurrent systems as well as distributed systems.
While Lamport diagrams can effectively visualize shared-memory systems as well as distributed ones, Lamport's formal executions are not well suited for the shared-memory domain, since processes and messages are often not the right abstractions.
With CSDs, we have a diagrammatic syntax \emph{and} a formal model that fit both domains.

\section{Syntax and Semantics of Causal Separation Diagrams}
\label{sec:csds-defined}

In \Cref{sec:csds-informally} we discussed the intuitions behind causal separation diagrams (CSDs), and how they arise from Lamport diagrams.
In this section we give a formal treatment of CSDs as terms of an inductive data type, and develop a concept of semantic interpretations of CSDs that we will make heavy use of in later sections.

\subsection{Site Configurations}
\label{sec:configurations}

Recall from \Cref{sec:csds-informally} that Lamport diagrams can be decomposed into a sequence of \textit{global steps}, where each adjacent pair of steps meets at a collection of sites called a \textit{site configuration} (or just \textit{configuration}).
The configuration at the start of a global step describes the state of the sites before that step takes place, while the configuration at the end describes the state of the sites after the step.
The diagram as a whole also starts and ends on a pair of configurations --- namely, the starting configuration of its first step, and the ending configuration of its last step.
A formally-defined CSD will have type \(\Gamma_1 \rightrightarrows \Gamma_2\), where \(\Gamma_1\) and \(\Gamma_2\) are \emph{bounding configurations} --- the configurations the diagram begins and ends on, respectively.
Site configurations are themselves terms, so \(\Gamma_1 \rightrightarrows \Gamma_2\) will be a \textit{dependent type}.
(In fact, nearly \emph{every} type we define will be dependent.)

\begin{definition}[Site configurations]
  \label{def:site-configurations}
  Let \(\tau\) be a universe of types with products. Then a \emph{site configuration} \(\Gamma\) is a binary tree with leaves drawn from \(\tau\), i.e., a term of the following grammar:
  \begin{alignat*}{3}
      &\Gamma && \coloneqq \Gamma \otimes \Gamma &&\;\mid\; [\tau] \\
      &\tau   && \coloneqq \tau \times \tau &&\;\mid\; \dots
  \end{alignat*}
\end{definition}

The leaf constructor \([-]\) gives the type of some state that is isolated at one site, while the spatial product \(\otimes\) models a kind of separating conjunction
\footnote{
  \textit{Separating conjunction} is a logical connective found in separation logic, where two properties of heaps can be conjoined if a heap can be split into two factors, one of which satisfies one property and one of which satisfies the other.
  A site configuration can thus be thought of as a particular factorization of a distributed heap.
}%
, giving the type of state that is spatially distributed over multiple sites.
For instance, if the type universe \(\tau\) includes naturals~\(\mathbb{N}\) and booleans~\(\mathbb{B}\), then \([\mathbb{N} \times \mathbb{B}] \otimes [\mathbb{B}]\) is a configuration with two sites, one carrying a pair of a natural and a boolean, and the other carrying a single boolean.

The spatial product \(\otimes\) is like a ``lifted'' version of the local product \(\times\); and like the local product, we will wish to treat \(\otimes\) as associative and commutative.
Since reordered/rebalanced binary trees are syntactically distinct terms, however, we introduce a type of permutations \(\sigma : \Gamma_1 \simeq \Gamma_2\) to mediate between equivalent configurations.

\begin{definition}[Sites]
  \label{def:site-type}
  The type \(\mathrm{Site}(\Gamma)\), defined recursively over the structure of configuration \(\Gamma\), is the type of paths from the root of \(\Gamma\) to each of its leaves:
  \begin{alignat*}{2}
    &\mathrm{Site}([\tau])
      &&= \top \\
    &\mathrm{Site}(\Gamma_1 \otimes \Gamma_2)
      &&= \mathrm{Site}(\Gamma_1) + \mathrm{Site}(\Gamma_2)
  \end{alignat*}
\end{definition}
\begin{definition}[Permutations of sites (\(\simeq\))]
  \label{def:site-permutations}
  The type of \emph{permutations} \(\Gamma_1 \simeq \Gamma_2\) is an equivalence relation on site configurations, defined so that its elements \(\sigma\) correspond to type-preserving bijections \(\mathrm{Site}(\Gamma_1) \rightarrow \mathrm{Site}(\Gamma_2)\).
  By abuse of notation, we denote by \(\sigma\) (and \(\sigma^{-1}\)) the bijection witnessed by \(\sigma\).
\end{definition}

In \Cref{def:site-type}, \(\top\) is the unit type (with single value \(\bullet\)), and \(+\) gives sum types (with injections \(\mathbf{inj}_\ell\) and \(\mathbf{inj}_r\)).
For example, the type of sites for \(([\mathbb{N}] \otimes [\mathbb{B}]) \otimes [\mathbb{B}]\) is \((\top + \top) + \top\).
To address the site of type \(\mathbb{N}\), we write the term \(\mathbf{inj}_\ell(\mathbf{inj}_\ell(\bullet))\), which tells us we can isolate this site by focusing along the left-hand subtrees of this configuration.

\subsection{Causal Separation Diagrams}
\label{sec:constructing-csds}

From \Cref{sec:csds-informally}, we know that CSDs have two forms of composition: sequential composition and concurrent composition.
\footnote{Some readers will recognize the syntax of CSDs as a (free) symmetric monoidal category. We will have more to say about categorical connections in \Cref{sec:discussion}; for now, we acknowledge the connections but proceed concretely.}
Just as conjunctive normal form makes Boolean formulae easier to work with, we will restrict concurrent composition to appear only under sequential composition.
Every CSD, then, has two layers: an outer list modeling sequencing, and an inner tree modeling concurrency.
To separate these layers, we give them distinct symbols: a diagram \(x : \Gamma_1 \rightrightarrows \Gamma_2\) is a diagram proper, and can be composed sequentially, while a diagram \(x : \Gamma_1 \multimap \Gamma_2\) is a global step, and can be composed concurrently.
These are morally both diagrams --- a global step is just a diagram in the process of being built --- and we will generally not distinguish between them.

\begin{definition}[Causal separation diagrams (\(\rightrightarrows\))]
  \label{def:causal-separation-diagram}
  A \emph{causal separation diagram} is a sequence of global steps (see \Cref{def:global-step}, next), constructed according to the following rules:\\

  \begin{center}
%

\bottomAlignProof
\AxiomC{}
\UnaryInfC{\(\mathbf{id} : \Gamma \rightrightarrows \Gamma\)}
\DisplayProof
\qquad
\bottomAlignProof
\AxiomC{\(x : \Gamma_1 \rightrightarrows \Gamma_2\)}
\AxiomC{\(y : \Gamma_2 \multimap \Gamma_3\)}
\BinaryInfC{\((x \mathop{;} y) : \Gamma_1 \rightrightarrows \Gamma_3\)}
\DisplayProof


  \end{center}
\end{definition}
\noindent
The \textbf{id} and sequencing (\textbf{;}) constructors play the same roles, respectively, as ``nil'' and ``cons'' do for inductive lists.
We take our sequences to grow to the right (a ``snoc'' list) from an initial \textbf{id} seed, and moreover require that adjacent global steps be compatible: if a step ends on one configuration, the following step must begin on the same configuration.

\begin{definition}[Global steps (\(\multimap\))]
  \label{def:global-step}
  A \emph{global step} is a binary tree of \emph{atomic steps}, constructed according to the rules below:\\

  \begin{center}
\qquad\qquad\qquad\quad
\bottomAlignProof
\AxiomC{\(x : \Gamma_1  \multimap \Gamma_2\)}
\AxiomC{\(y : \Gamma_1' \multimap \Gamma_2'\)}
\BinaryInfC{\((x \parallel y) : \Gamma_1 \otimes \Gamma_1' \multimap \Gamma_2 \otimes \Gamma_2'\)}
\DisplayProof
\hfill
\bottomAlignProof
\AxiomC{}
\UnaryInfC{\(\mathbf{fork} : [\tau \times \tau'] \multimap [\tau] \otimes [\tau']\)}
\DisplayProof
\qquad\qquad\qquad

\qquad

\qquad\qquad
\bottomAlignProof
\AxiomC{\(\sigma : \Gamma_1 \simeq \Gamma_2\)}
\UnaryInfC{\(\mathbf{perm}\ \sigma : \Gamma_1 \multimap \Gamma_2\)}
\DisplayProof
\hfill
\bottomAlignProof
\AxiomC{}
\UnaryInfC{\(\mathbf{tick} : [\tau_1] \multimap [\tau_2]\)}
\DisplayProof
\hfill
\bottomAlignProof
\AxiomC{}
\UnaryInfC{\(\mathbf{join} : [\tau] \otimes [\tau'] \multimap [\tau \times \tau']\)}
\DisplayProof
\qquad\qquad\qquad

  \end{center}
\end{definition}
The atomic steps \(\mathbf{tick}\), \(\mathbf{fork}\), \(\mathbf{join}\), and \(\mathbf{perm}\) describe the elementary ways in which sites can be transformed over time.
The concurrence (\(\parallel\)) operator fuses two global steps into one.
Since the two steps must operate over distinct configurations, no atomic step can share a site with any concurrent step.
Thus, just as \(\otimes\) acts like a separating conjunction, \(\parallel\) acts like the concurrency rule of concurrent separation logic.
(We discuss future work following this analogy in \Cref{sec:discussion}.)

The \textbf{perm} constructor transforms a configuration into any equivalent configuration according to the type of permutations \(\simeq\) of \Cref{def:site-permutations}.
It will be convenient to have shorthand for three special cases of \textbf{perm}:
\begin{itemize}
  \item \(\mathbf{noop} : \Gamma \multimap \Gamma\) is a step over the identity permutation;

  \item \(\mathbf{swap} : [\tau] \otimes [\tau'] \multimap [\tau'] \otimes [\tau]\) is a step commuting two sites; and

  \item \(\mathbf{assoc} : \Gamma_1 \otimes (\Gamma_2 \otimes \Gamma_3) \multimap (\Gamma_1 \otimes \Gamma_2) \otimes \Gamma_3\) is a step reassociating a configuration.
\end{itemize}

The \textbf{tick} constructor models any arbitrary local transformation of state.
For instance, a \(\mathbf{tick}\) of type \([\mathbb{N}] \multimap [\mathbb{N} \times \mathbb{B}]\) might describe an action which prepares a (boolean) message depending on the current (numeric) state.
We deliberately leave the local transformations unconstrained to avoid parameterizing CSDs over yet another type.
Concrete information about each individual \(\mathbf{tick}\) can instead be associated to a CSD by way of \emph{labeling}, which we will discuss in \Cref{sec:labeled-csds}.

The \textbf{fork} and \textbf{join} constructors reify the connection between spatial and local products alluded to in \Cref{sec:configurations}.
If we have a local pair of state at one site --- for instance, a pair \([\mathbb{N} \times \mathbb{B}]\) of numeric state and prepared message --- we can spatially separate its components onto two sites with \textbf{fork}.
Conversely, state distributed over two sites can be fused into a local product on one site with \textbf{join}.
Therefore, these steps are our analogues of the send/receive actions found in Lamport executions.

Although a traditional Lamport diagram treats send and receive actions as state-modifying actions, we factor them into two separate steps: a Lamport-style send is realized as a \textbf{tick} followed by a \textbf{fork}, and a Lamport-style receive is realized as a \textbf{join} followed by a \textbf{tick}.
\footnote{To obtain a legitimate CSD from \Cref{fig:diagram-decompositions}(b), we would need to extract the implicit \(\mathbf{tick}\) from each send and receive action.}
This factorization allows us to treat \emph{all} modifications of local state uniformly via \(\mathbf{tick}\), which helps us greatly when associating concrete operations to each \(\mathbf{tick}\)~(\Cref{sec:labeled-csds}).

\begin{figure}%
  \centering%
  \begin{minipage}{0.3\columnwidth}%
    \centering%
    \includegraphics[width=0.6\columnwidth]{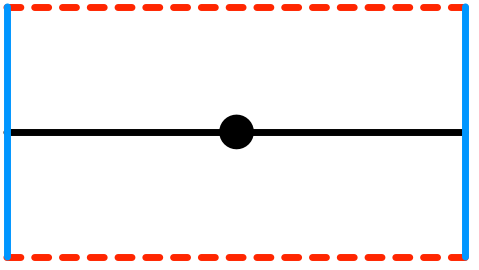}\\
    \textbf{tick}%
  \end{minipage}%
  \begin{minipage}{0.3\columnwidth}%
    \centering%
    \includegraphics[width=0.6\columnwidth]{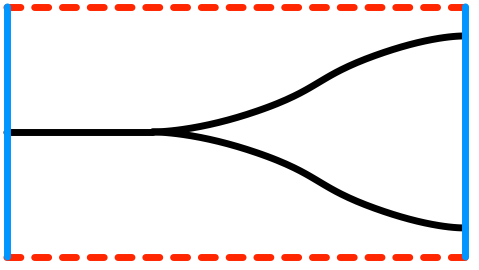}\\
    \textbf{fork}%
  \end{minipage}%
  \begin{minipage}{0.3\columnwidth}%
    \centering%
    \includegraphics[width=0.6\columnwidth]{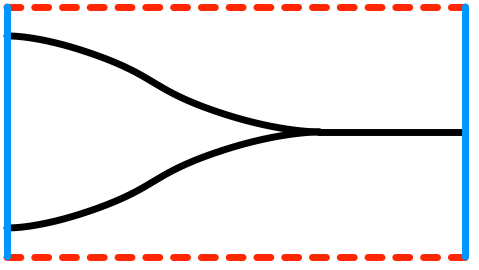}\\
    \textbf{join}%
  \end{minipage}\\
  \vspace{1em}
  \begin{minipage}{0.3\columnwidth}%
    \centering%
    \includegraphics[width=0.6\columnwidth]{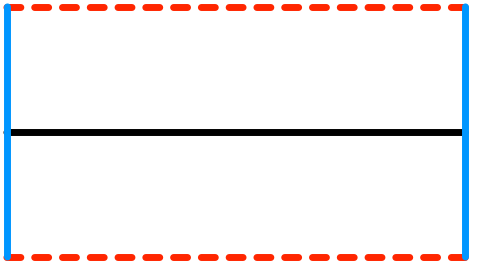}\\
    \textbf{noop}%
  \end{minipage}%
  \begin{minipage}{0.3\columnwidth}%
    \centering%
    \includegraphics[width=0.6\columnwidth]{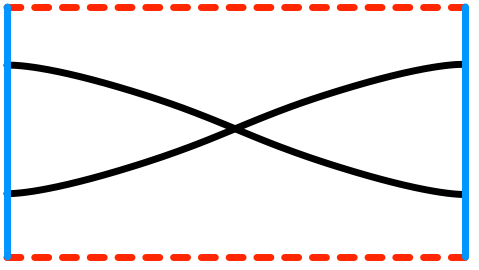}\\
    \textbf{swap}%
  \end{minipage}%
  \begin{minipage}{0.3\columnwidth}%
    \centering%
    \includegraphics[width=0.6\columnwidth]{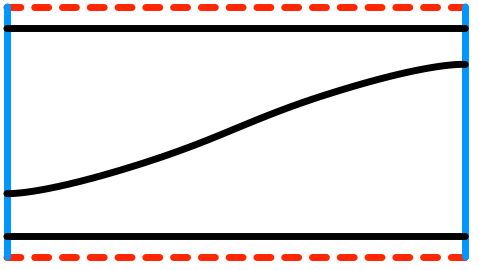}\\
    \textbf{assoc}%
  \end{minipage}
  \caption{Atomic steps of a CSD, depicted as graphical tiles. The \(\mathbf{noop}\), \(\mathbf{swap}\), and \(\mathbf{assoc}\) tiles characterize the more general \(\mathbf{perm}\) atomic step.}
  \label{fig:csd-tiles}
\end{figure}

\Cref{fig:csd-tiles} depicts the \textbf{tick}, \textbf{fork}, \textbf{join}, \textbf{noop}, \textbf{swap}, and \textbf{assoc} atomic steps graphically.
These tiles can be freely composed along like boundaries (that is, solid blue lines compose with solid blue lines, and dashed red lines compose with dashed red lines) to construct whole diagrams, so long as any sequenced pair of diagrams agree on the arrangement of sites crossing between them.
For instance, consider the CSD given by the term
\(
  \mathbf{id} \mathbin{;}
  (\mathbf{tick} \parallel \mathbf{fork}) \mathbin{;}
  \mathbf{assoc} \mathbin{;}
  (\mathbf{join} \parallel \mathbf{noop})
\).
As a (snoc)-list, this CSD begins from an empty diagram (\(\mathbf{id}\)) to which successive global steps are appended (with~\(\mathbin{;}\)).
Each constituent global step is built up as a concurrent composition of atomic steps (with~\(\parallel\)).
We can better display the structure of this CSD diagrammatically:
\begin{figure}[h]%
  \begin{center}%
    \includegraphics[width=0.4\columnwidth]{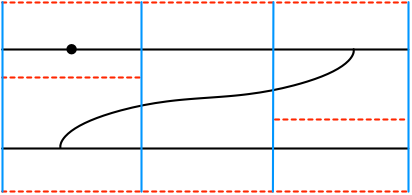}%
  \end{center}%
\end{figure}%

\noindent
We begin on some site configuration \([\tau_1] \otimes [\tau_2 \times \tau_3]\), and perform a \textbf{tick} on the first site and a \textbf{fork} on the second site to reach configuration \([\tau'_1] \otimes ([\tau_2] \otimes [\tau_3])\), where \(\tau'_1\) is the result type of the \textbf{tick}.
With \textbf{assoc}, we then rebalance the configuration into \(([\tau'_1] \otimes [\tau_2]) \otimes [\tau_3]\), so that the following step can \textbf{join} the first two sites (while leaving the third alone with \textbf{noop}).
This CSD thus ends on configuration \([\tau'_1 \times \tau_2] \otimes [\tau_3]\).
Since the type \(\tau_2\) ends up migrating from one site to another, this CSD might describe a message sent from one process to another.

\paragraph{Abuses of notation}
Since CSDs are lists of global steps, we can define a version of concurrent composition that acts over entire CSDs by zipping them together (with \textbf{noop} padding if their lengths are mismatched) and composing each pair.
Likewise, we can sequentially extend a CSD by another CSD using the equivalent of a \textit{concat} operator.
Rather than allocate new symbols to these binary operators, we will abuse notation, letting \(\parallel\) and \(\mathbin{;}\) stand in for them.

In our Agda mechanization, the indexed types \(\rightrightarrows\) and \(\multimap\) are unified in a type with an auxiliary index over \(\{\mathrm{Seq}, \mathrm{Par}\}\).
Throughout the rest of this paper, we take advantage of this technical contrivance to define single functions that can pattern-match through both sequential and concurrent layers of a CSD, instead of defining a separate function for each layer.

\subsection{Labeled CSDs}
\label{sec:labeled-csds}

Recall that a \textbf{tick} step is meant to model a local transformation of state.
However, up to this point, there is no way to specify \textit{what} that local transformation actually is for each \textbf{tick}.
If we only have one transformation in a given setting, we can interpret each tick as that specific transformation.
But this is clearly too much of a limitation --- most systems can do more than one thing!

While we could parameterize CSDs over a type of actions (and construct each \textbf{tick} with a choice of action), this would complicate the type signature of CSDs, and introduce data for which the CSD itself is simply a carrier.
Instead, we follow the pattern of \textit{container types}~\citep{altenkirch-containers}, in which the places where data can be held are characterized separately from the assignment of data to those places.
For example, the generic type of lists \(\mathrm{List}(T)\) can be factored into two parts: a Peano natural \(n : \mathbb{N}\) and an assignment \(\mathrm{Fin}(n) \rightarrow T\) of values to indices.
The Peano natural \(n\) describes a particular \emph{shape} of list (with zero playing the role of the empty list, and the successor constructor playing the role of list consing), while \(\mathrm{Fin}(n)\) characterizes the positions within a list of that shape.
The assignment \(\mathrm{Fin}(n) \rightarrow T\) then fills those positions with concrete values.

\begin{definition}[The type of ticks]
  \label{def:tick-type}
  For a CSD \(X\), the type \(\mathrm{Tick}(X)\) has precisely one value for every \textbf{tick} in \(X\), and is defined recursively over the structure of \(X\):
  \begin{alignat*}{5}
    &   \mathrm{Tick}(\mathbf{tick})
    &&= \top
  \\
    &   \mathrm{Tick}(\mathbf{fork})
    &&= \bot
  \\
    &   \mathrm{Tick}(\mathbf{join})
    &&= \bot
  \\
    &   \mathrm{Tick}(\mathbf{perm}\ \sigma)
    &&= \bot
  &&\qquad
    &&  \mathrm{Tick}(\mathbf{id})
    &&= \bot
  \\
    &   \mathrm{Tick}(x \parallel y)
    &&= \mathrm{Tick}(x) + \mathrm{Tick}(y)
  &&\qquad
    &&\mathrm{Tick}(x \mathbin{;} y)
    &&= \mathrm{Tick}(x) + \mathrm{Tick}(y)
  \end{alignat*}
  Here, \(\bot\) is the empty type, \(\top\) is the unit type (with only value \(\bullet\)), and \(+\) gives sum types (with injections \(\mathbf{inj}_\ell\), \(\mathbf{inj}_r\)).
\end{definition}

\begin{definition}[Labeled CSDs]
  \label{def:labeled-csds}
  A \emph{\(T\)-labeling} \(f : \mathrm{Tick}(X) \rightarrow T\) assigns a value of type \(T\) to every \textbf{tick} in \(X\).
  A \emph{\textbf{\(T\)-labeled CSD}}, written \(\langle X , f \rangle : \Gamma_1 \rightrightarrows^T \Gamma_2\), is a diagram together with a \(T\)-labeling.
\end{definition}

Given a labeled CSD, we can restrict its labeling to a subdiagram by pre-composing with the left or right injection for sums.
For instance, the prefix of the labeled CSD \(\langle (x \mathbin{;} y), f \rangle\) can be obtained as \(\langle x, f \circ \mathbf{inj}_\ell \rangle\).
In the base case, we end up with \(\langle \mathbf{tick}, \bullet \mapsto v \rangle\) --- precisely a \textbf{tick} annotated with a value.
This makes labeled CSDs an excellent solution for specifying the behavior of each \textbf{tick}.

In a traditional execution (\Cref{def:lamport-execution}), every local action comes with some information built in --- not what the action is, but \emph{who} performed it.
This is because every action occurs on a particular process's total order.
Although CSDs do not treat process lines specially, we can include this same information by positing a type \(\mathrm{Pid}\) of process identifiers, and working in terms of \(\mathrm{Pid}\)-labeled CSDs.

\subsection{Semantic Interpretations of CSDs}
\label{sec:interpreting-csds}

The construction of the \(\mathrm{Tick}\) type in \Cref{def:tick-type} is our first example of an \emph{interpretation} of CSDs: we assigned some type to each atomic step, and described how sequential and concurrent composition act over those types to yield a type for larger diagrams.
This pattern is emblematic of denotational semantics: ``the meaning of the composition is the composition of the meanings.''
\footnote{This compositionality principle appears to be folklore in denotational semantics; we cannot find a canonical source. It dates at least to Frege, in the context of natural languages.}
By itself, the CSD representation is not much use; its utility comes from its interpretability.

\begin{definition}[Semantic interpretations]
  \label{def:interpretations}
  A \emph{semantic interpretation} (or \emph{semantics}, or \emph{interpretation}) of CSDs is a function \((\Gamma_1 \rightrightarrows \Gamma_2) \rightarrow F(\Gamma_1, \Gamma_2)\) mapping each CSD to a semantic domain \(F\) indexed by site configurations.
  \footnote{The domain \(F\) ought to be a symmetric monoidal category, with an interpretation being a functor from \(\rightrightarrows\) to \(F\). However, we neither prove nor require that \(\rightrightarrows\) be such a category --- although we are eager to make such connections in the future.}
\end{definition}

In the case of \(\mathrm{Tick}\), we take \(F(-,-)\) to be the universe of types, \(\mathbf{Type}\), without dependence on the particular bounding configurations.
Much of the rest of this paper will be devoted to the construction and analysis of additional interpretations, following the landmarks given in the introduction:
\begin{itemize}
  \item
    In \Cref{sec:interp-causal-paths}, we give a semantics in \(F(\Gamma_1, \Gamma_2) = \mathrm{Site}(\Gamma_1) \rightarrow \mathrm{Site}(\Gamma_2) \rightarrow \mathbf{Type}\), a domain of types \(\rightsquigarrow\) whose elements \(p_{12} : s_1 \rightsquigarrow s_2\) are \emph{causal paths} between sites at the boundaries of the diagram.
    This yields a proof-relevant analogue of \citeauthor{lamport-clocks}'s \happensbefore{} relation, where a path gives concrete evidence for why its endpoints are causally related.

  \item
    In \Cref{sec:interp-clock-programs}, we give a semantics in \(F(\Gamma_1, \Gamma_2) = \mathrm{Valuation}(\Gamma_1) \rightarrow \mathrm{Valuation}(\Gamma_2)\), a domain of functions \(\mathcal{C}\), parametric in a choice of logical clock.
    A valuation \(\nu : \mathrm{Valuation}(\Gamma_1)\) is an assignment \(\mathrm{Site}(\Gamma_1) \rightarrow \mathrm{Time}\) of timestamps to each site; so functions \(\mathcal{C}\) compute timestamps \(\mathcal{C}_\nu\) on \(\Gamma_2\) from timestamps \(\nu\) on \(\Gamma_1\).

  \item
    In \Cref{sec:interp-proofs}, we give a semantics in \(F(\Gamma_1, \Gamma_2) = \forall s_1\ s_2.\ (s_1 \rightsquigarrow s_2) \rightarrow (\forall \nu.\ \nu(s_1) \le \mathcal{C}_\nu(s_2))\), a domain of \emph{proofs} relating the first two interpretations via Lamport's clock condition.
    \footnote{Although it looks like \(\Gamma_1\) and \(\Gamma_2\) are not used in this domain, we are using the \(\rightsquigarrow\) and \(\mathcal{C}\) obtained from the other two interpretations, which very much do depend on the given configurations.}
    The resulting proof is constructed modularly, by composing proofs over atomic steps into proofs over whole diagrams, and is parametric in a choice of logical clock.
\end{itemize}

Our target domains (\happensbefore{}, logical clocks, and the clock condition) are all pre-existing concepts in the literature.
However, the interpretations sketched above only directly relate points on the beginning and ending \emph{boundaries} of a diagram, while these concepts traditionally speak of points \emph{interior} to a diagram.
To bridge this gap, we provide a general, two-phase recipe for building interpretations.

\begin{itemize}
  \item
    First we define a ``spanning'' interpretation, restricting the target domain to relationships between the initial and final sites of a CSD.
    These interpretations are typically easy to implement recursively over the structure of a CSD.
    For the causal paths of \Cref{sec:interp-causal-paths}, this will yield a domain of ``spanning paths'' giving causal relationships only between the sites on the boundary of a diagram.

  \item
    Next we define an ``interior'' interpretation, extending the first interpretation to include relationships between points on the interior of a diagram \(X\).
    For causal paths, an ``interior path'' will be a spanning path across any subdiagram of \(X\), so our interpretation will relate sites in any of the site configurations visited by \(X\).
\end{itemize}

The interpretations presented in \Cref{sec:interp-causal-paths,sec:interp-clock-programs,sec:interp-proofs} all follow this same recipe.

\section{The Inductive Type of Causal Paths}
\label{sec:interp-causal-paths}

In this section we develop a notion of causal order within CSDs that captures the potential flows of information through a concurrent system.
These flows are traditionally visualized in Lamport diagrams as geometric paths, reducing causality to a kind of connectivity between two points in space and time.
We take these paths seriously as \textit{bona fide} data: the type of \emph{causal paths} is defined by a semantic interpretation of CSDs, following the pattern established in \Cref{sec:interpreting-csds}.
This results in a causal relation that is \emph{proof-relevant}: rather than the mere fact that ``\(e_1\) happens before \(e_2\)'' observed in traditional executions, we have concrete (and potentially multiple) paths \(p : e_1 \rightsquigarrow e_2\).
Such witnesses become extremely useful in proof by induction, including those we present in \Cref{sec:interp-proofs} for logical clocks.

\subsection{Spanning Paths}
\label{subsec:spanning-paths}

We first restrict our attention to causal relationships between sites in the bounding configurations of a diagram, which we will hereafter call \emph{bounding sites}.
In \Cref{subsec:interior-paths}, we will extend these relationships to sites on any configuration visited by a diagram.

\begin{definition}[Spanning relations]
  \label{def:spanning-relations}
  A \emph{spanning relation} between configurations \(\Gamma_1, \Gamma_2\) is a type family \(\mathrm{Site}(\Gamma_1) \rightarrow \mathrm{Site}(\Gamma_2) \rightarrow \mathbf{Type}\) taking a pair of sites to a type of relationships between them.
\end{definition}

If \(\rightsquigarrow\) is a spanning relation, an element of type \(s_1 \rightsquigarrow s_2\) describes a potential flow of information between sites \(s_1\) and \(s_2\).
Because information might take one of many branching and converging paths \emph{en route} between any pair of sites, \(s_1 \rightsquigarrow s_2\) may have multiple distinct values.
This makes spanning relations \emph{proof-relevant}: knowing that \(s_1 \rightsquigarrow s_2\) means knowing \emph{why} that fact is true.

Given two spanning relations \(\rightsquigarrow_1\) and \(\rightsquigarrow_2\), we can compose them sequentially or concurrently.
Sequential composition is standard relational composition (\(\circ\)): we have a path across the sequence of two spanning relations if we have paths across each individually that meet at some common site.
Concurrent composition is a disjoint sum (\(+\)): we have a path across the concurrence of two spanning relations if we have a path across either individually.

Every CSD induces a spanning relation modeling the concrete ways information can flow from one side of the diagram to the other.
These are precisely the paths that the Lamport diagram makes evident graphically.

\begin{definition}[Spanning paths]
  \label{def:spanning-paths}
  The type family \(\mathrm{Span}(X)\) of \emph{spanning paths} through a CSD \(X : \Gamma_1 \rightrightarrows \Gamma_2\) is a spanning relation, and is defined inductively over the structure of \(X\):
  \begin{alignat*}{5}
    &   \mathrm{Span}(\mathbf{tick})
    &&= \lambda s_1\ s_2.\ \top \\
    &   \mathrm{Span}(\mathbf{fork})
    &&= \lambda s_1\ s_2.\ \top \\
    &   \mathrm{Span}(\mathbf{join})
    &&= \lambda s_1\ s_2.\ \top \\
    &   \mathrm{Span}(\mathbf{perm}\ \sigma)
    &&= \lambda s_1\ s_2.\ (s_2 \equiv \sigma(s_1))
  &&\qquad
    &&  \mathrm{Span}(\mathbf{id})
    &&= \lambda s_1\ s_2.\ (s_2 \equiv s_1) \\
    &   \mathrm{Span}(x \parallel y)
    &&= \mathrm{Span}(x) + \mathrm{Span}(y)
  &&\qquad
    &&  \mathrm{Span}(x \mathbin{;} y)
    &&= \mathrm{Span}(y) \circ \mathrm{Span}(x)
  \end{alignat*}
  When \(X\) is understood, we write \(s_1 \rightsquigarrow s_2\) to mean \(\mathrm{Span}(X)(s_1, s_2)\).
\end{definition}

The \(\mathbf{tick}\), \(\mathbf{fork}\), and \(\mathbf{join}\) steps are interpreted trivially into the unit type \(\top\), because those steps have precisely one path for every opposing pair of bounding sites: \(\mathbf{join}\), for instance, relates two input sites to one output site, and information on both inputs will flow into the single output.
Meanwhile, \(\mathbf{id}\) relates a configuration to itself (so only matching indices are connected by paths); and \(\mathbf{perm}\ \sigma\) relates inputs to outputs according to the permutation of sites performed by \(\sigma\).

\begin{figure}
  \begin{center}
    \includegraphics[width=0.7\columnwidth]{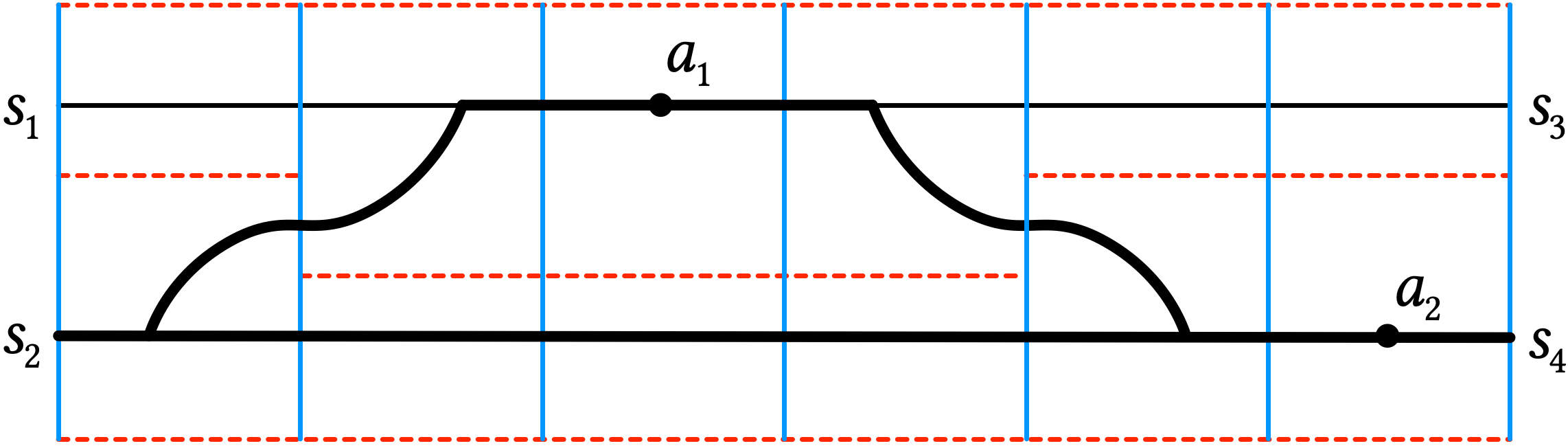}
  \end{center}
  \caption{In this diagram, the bolded paths identify distinct witnesses to the causal relationship between initial site \(s_2\) and final site \(s_4\).}
  \label{fig:multiple-paths}
\end{figure}
For example, the CSD depicted in \Cref{fig:multiple-paths} goes from configuration \(s_1 \otimes s_2\) to configuration \(s_3 \otimes s_4\).
Because \(s_2\) is causally related to \(s_4\) by two distinct paths, the type \(s_2 \rightsquigarrow s_4\) has two inhabitants.

\subsection{Interior Paths}
\label{subsec:interior-paths}

Next, we will extend our spanning relation between bounding sites to a relation on \emph{all} points of interest within a diagram.
To do this, we need to refer not only to sites in the bounding configurations of \(X\), but on \emph{any} site configuration visited by \(X\).
A CSD with a sequence of \(N\) global steps visits \(N+1\) site configurations: one at the start of the diagram, and one at the end of each global step.
Hence, an \emph{event} will be a choice of site configuration in a diagram, together with a choice of site within that configuration.

\begin{definition}[Cuts]
  \label{def:cuts}
  The type \(\mathrm{Cut}(X)\) of \emph{cuts} within a diagram \(X : \Gamma_1 \rightrightarrows \Gamma_2\) has one inhabitant for every site configuration visited by \(X\), and is defined recursively over the structure of \(X\).
  The associated function \(\mathit{cut}(-)\) picks out the site configuration for each index of \(\mathrm{Cut}(X)\).
  \begin{alignat*}{8}
    &   \mathrm{Cut}(\mathbf{id})
    &&= \top
  &&\qquad
    &&
    &&
  &&\quad
    &&  \mathit{cut}(\bullet)
    &&= \Gamma_2
  \\
    &   \mathrm{Cut}(x \mathbin{;} y)
    &&= \mathrm{Cut}(x) + \top
  &&\qquad
    &&  \mathit{cut}(\mathbf{inj}_\ell(t))
    &&= \mathit{cut}(t)
  &&\quad
    &&  \mathit{cut}(\mathbf{inj}_r(\bullet))
    &&= \Gamma_2
  \end{alignat*}
\end{definition}

\begin{definition}[Events]
  \label{def:events}
  The type \(\mathrm{Event}(X)\) of \emph{events} in a diagram \(X\) is the type of points in spacetime consisting of a temporal coordinate (a cut) together with a spatial coordinate (a site):
  \begin{alignat*}{2}
    &\mathrm{Event}(X)
      &&= (t : \mathrm{Cut}(X),\ s : \mathrm{Site}(\mathit{cut}(t)))
  \end{alignat*}
\end{definition}

This order of coordinates inverts the convention for events in a traditional execution, where we first select a process (a spatial coordinate) and then select an action occurring on that process (a temporal coordinate).
In our figures (such as \Cref{fig:multiple-paths}), events exist wherever a line modeling the flow of data (in black) intersects a consistent cut (in blue).

Care should be taken not to confuse \emph{events} with \emph{actions}.
In the traditional model of executions, an ``event'' is modeled by a local action --- the equivalent of our \(\mathbf{tick}\).
However, since an action is effectively a discontinuous, instantaneous change to state, this leads to questions about what the state of a system is ``at'' a local action:
Has the action actually happened yet or not?
Is the action included in its own causal history?
These ties are usually broken by interpreting events to occur either slightly before or slightly after an action --- and sometimes both, depending on context.
We prefer not to conflate these concepts in the first place: for us, an event is no more than a point in space at a point in time, with no presumption that it is special in any particular way.

Next, we need a way to describe paths between any two events.
For any two cuts in a CSD, we can consider the global steps between them as a subdiagram.
Then a path between two events is no more than a path spanning the subdiagram between their cuts.
Order matters, however: if a CSD passes through distinct cuts \(t_1, t_2\) (in that order), the subdiagram ``from \(t_2\) to \(t_1\)'' does not really exist --- at least not in the expected sense.
To preclude such inversions, we will define subdiagrams only over legal intervals.

\begin{definition}[Intervals]
  \label{def:intervals}
  The \emph{interval} \(t_1 \cdots t_2\) between cuts \(t_1, t_2\) in a diagram \(X\) is the type with a (unique) inhabitant \(t_{12}\) if and only if \(X\) visits \(t_1\) no later than \(t_2\).
\end{definition}

\begin{definition}[The subdiagram over an interval]
  \label{def:subdiagrams-during}
  The \emph{subdiagram over an interval \(t_{12} : t_1 \cdots t_2\)}, denoted \(\mathit{during}(t_{12})\), is the CSD consisting of the global steps appearing strictly between cuts \(t_1, t_2\) in a diagram \(X\).
\end{definition}

Since CSDs are effectively (snoc-)lists at the top level, using \(\mathit{during}(-)\) is akin to using the common list functions \verb|drop| and \verb|take|: we drop everything after both cuts, then take everything that remains after the first cut.

Finally, we can obtain a causal relation between events:

\begin{definition}[Causal relations]
  \label{def:causal-relations}
  For a diagram \(X\), a \emph{causal relation} is a type family \(\mathrm{Event}(X) \rightarrow \mathrm{Event}(X) \rightarrow \mathbf{Type}\) taking every pair of events to a type of relationships between them.
\end{definition}

\begin{definition}[Causal paths]
  \label{def:causal-paths}
  The type family \(\rightsquigarrow\) of \emph{causal paths} (sometimes \emph{interior paths}) through a diagram \(X\) is a causal relation.
  The inhabitants of \(e_1 \rightsquigarrow e_2\) are (dependent) pairs consisting of an interval between the events together with a spanning path under that interval:
  \[(t_1, s_1) \rightsquigarrow (t_2, s_2) = (t_{12} : t_1 \cdots t_2,\ p_{12} : \mathrm{Span}(\mathit{during}(t_{12}))(s_1, s_2))\]
\end{definition}

We consistently pun \(\rightsquigarrow\) to mean either spanning paths or causal paths depending on whether its arguments are sites or events.
Similar liberties will be taken (and acknowledged) with the interpretations of \Cref{sec:interp-clock-programs,sec:interp-proofs}.

The causal relation \(\rightsquigarrow\) enjoys reflexivity
\footnote{Unlike Lamport's \happensbefore{}, our \(\rightsquigarrow\) is reflexive. Since reflexive and irreflexive partial orders are in one-to-one correspondence, the choice largely comes down to a matter of preference.}%
, antisymmetry, and transitivity, making it a partial order.
As a proof-relevant type, reflexivity arises from the existence of unit paths, and transitivity arises from the composition of paths --- which is, moreover, strictly associative.
Unlike traditional executions (\Cref{def:lamport-execution}), antisymmetry is guaranteed by construction for every CSD: it is impossible to introduce a causal loop because state flows only forward in time.
Proofs of these properties can be found in our Agda development; we elide them here for brevity.

\paragraph{An order on actions}
Here and in \Cref{sec:csds-informally}, we were careful to distinguish the actions related by \happensbefore{} from the spacetime coordinates we call events.
Nonetheless, the two notions are closely related: every local action \(a\) has a pair of associated events \(e_a^\ell, e_a^r\) before and after it.
We can use these events to act as proxy for the actions in our system to recover an irreflexive order on actions: \(a_i < a_j\) if and only if \(e_{a_i}^r \rightsquigarrow e_{a_j}^\ell\).
For example, in \Cref{fig:multiple-paths}, we have \(a_1 < a_2\), since \(e_{a_1}^r \rightsquigarrow e_{a_2}^\ell\).
Because of this correspondence, we speak only of events in what follows --- we can always choose a suitable event to stand in for any action of interest.

\section{Interpreting CSDs into Logical Clocks}
\label{sec:interp-clock-programs}

In this section (and \Cref{sec:interp-proofs,sec:verified-clocks}) we apply CSDs to the analysis of \emph{logical clocks}, a common class of devices for reifying causal information into a concurrent system at runtime.
As \citet{lamport-clocks} observed, we often cannot rely on physical timekeeping to coordinate agents in a concurrent system: one agent's clock may drift relative to the others, and messages may take variable (or unbounded) amounts of time to propagate from sender to recipient.
Logical clocks solve this problem by measuring time against the occurrence of intentional \emph{actions} of the agents in the system.

\begin{figure}
{%
\hfill%
(a)
\includegraphics[width=0.40\columnwidth]{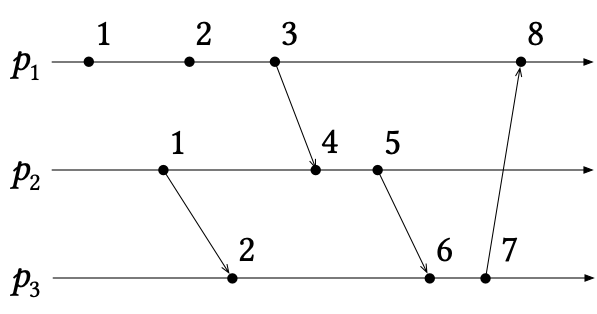}%
\qquad%
(b)
\includegraphics[width=0.40\columnwidth]{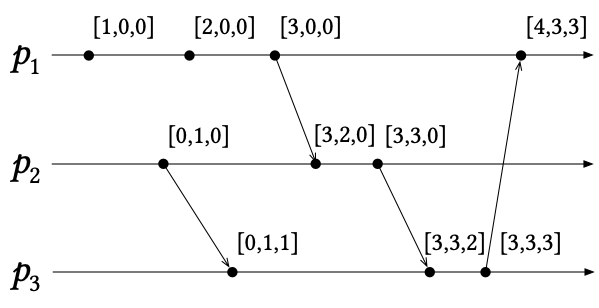}%
\hfill%
}
\caption{An example execution with an assignment of timestamps by the (a) Lamport clock and (b) vector clock.}
\label{fig:clock-examples}
\end{figure}

In the setting of \citet{lamport-clocks}, a \textbf{logical clock} (or just \emph{clock}) is a global assignment of partially-ordered values (called \emph{timestamps}) to actions in a concurrent execution.
\Cref{fig:clock-examples} gives examples of these assignments for two widely used logical clocks: the scalar clock~\citep{lamport-clocks} and the vector clock~\citep{mattern-vector-time, fidge-vector-time}, which respectively use scalar and vector timestamps.
We will discuss the specifics of these clocks in more detail in \Cref{sec:verified-clocks}, along with matrix clocks~\citep{wuu-bernstein-matrix-clocks,rst-matrix-clock}.

In our setting, a clock will assign a timestamp to every \emph{event} in a CSD.
Just as in \Cref{subsec:interior-paths}, we can assign timestamps to \emph{actions} by choosing an adjacent event to represent that action.

We will use a common formulation of clocks as implementations of an abstract data type with local \emph{increment} and \emph{merge} operations~\citep{raynal-singhal-survey}, and we bridge this local characterization of clocks into a global assignment of timestamps via interpretation.
We begin by justifying this choice of formulation; then, just as in the case of causal paths~(\Cref{sec:interp-causal-paths}), we construct an interpretation of CSDs \(X : \Gamma_1 \rightrightarrows \Gamma_2\) into a \emph{spanning} domain, in which an assignment of timestamps (or ``valuation'') on the sites of \(\Gamma_1\) is updated into a valuation on \(\Gamma_2\).
We conclude by extending this interpretation to an \emph{interior} domain, which will assign timestamps to all events within a diagram.

\subsection{Realizable Clocks}
\label{subsec:realizable-clocks}

In practical implementations, a logical clock is realized as a data structure, instantiated by each agent in a concurrent system, that tracks the passage of (logical) time from the perspective of that agent.
The timestamp associated to any action is that displayed by the clock of the agent when it performed the action.
The archetypal logical clock is the scalar clock of \citet{lamport-clocks}, in which every agent's clock maintains a single monotonically-increasing integer.
To ensure that every action occurs at a later ``time'' than those that occur causally prior, the scalar clock increments with each action, and updates to the maximum of its timestamp and that of any message received at that agent.
This property --- that causally-related actions have like-ordered timestamps --- is so important that it is called the \emph{clock condition}, and is required of \emph{any} prospective logical clock.
\footnote{Because our causal relation \(\rightsquigarrow\) is reflexive, our formulation of the clock condition does not guarantee that causally-related actions have distinct timestamps. We see this as a feature, not a bug: a clock need not tick for every local action, only those actually related to its purpose. Given knowledge about which actions are relevant, a strictly-increasing clock condition can be proved by the same methods of \Cref{sec:interp-proofs}.}

While we can always build a global assignment of timestamps from a system of clock replicas, we cannot always go in the reverse direction: a clock in the global sense may not be realizable as a data structure.
For instance, given an execution with \(n\) actions, if \(\mathcal{C}[-]\) is a monotone assignment of integer timestamps to this execution, then so is \(\mathcal{C}[-] + n\).
But an agent early in the execution has no knowledge of how many actions \emph{will occur} in total: any prediction it makes may be invalidated depending on what transpires in the future.
So even if \(\mathcal{C}[-]\) can be realized as a system of local clock instances, \(\mathcal{C}[-] + n\) certainly cannot be.

We restrict our attention to such \emph{realizable clocks}, as these make up the majority of clocks in the literature.
\footnote{Actually, we are not directly aware of any unrealizable clocks as such; though offline analyses of recorded execution traces might make good use of them.}
Following \citet{raynal-singhal-survey}, we treat logical clocks as an abstract data type (ADT) with two operators, \textit{increment} and \textit{merge}.
In addition, we assume a type \(\mathrm{Act}\) of actions performable by any agent in the system.

\begin{definition}[Clocks as an ADT]
  \label{def:logical-clocks}
  A \emph{logical clock} is a type \(\mathrm{Time}\) together with
  \begin{itemize}
    \item a family of operations \(\mathbf{increment}_a\) of type \(\mathrm{Time} \rightarrow \mathrm{Time}\) for every \(a : \mathrm{Act}\),

    \item an operation \(\mathbf{\sqcup}\) (pronounced \(\mathbf{merge}\)) of type \(\mathrm{Time} \times \mathrm{Time} \rightarrow \mathrm{Time}\).
  \end{itemize}

  Moreover, \(\mathrm{Time}\) must be preordered by a relation \(\le\), such that for all timestamps \(t_1, t_2 : \mathrm{Time}\), the above operations are inflationary:
  \begin{itemize}
    \item \(t_1 \le \mathbf{increment}_a(t_1)\),
    \item \(t_1 \le (t_1 \sqcup t_2)\), and
    \item \(t_2 \le (t_1 \sqcup t_2)\).
  \end{itemize}
\end{definition}

The \(\mathbf{increment_a}\) operation advances the clock's time depending on what the action \(a\) is.
For instance, a vector clock maintains an index for every agent, and it increments a \emph{different} index depending on which agent performed the action.
Since a CSD doesn't carry information about the provenance of an action, we take the elements of \(\mathrm{Act}\) to include that information themselves.
\footnote{Alternatively, we can take \(\mathrm{Act}\) to be the type of process identifiers, so that any agent may increment any index of the clock --- even one not intended to track that agent. \Cref{subsec:verifying-classifier-clocks} develops this perspective in more depth.}

The \(\mathbf{merge}\) operation advances the clock's time to any time after the two given timestamps.
This operation is used when an agent receives a message decorated with the sender's timestamp: by merging the sender's timestamp with the recipient's timestamp, any action occuring from that point on is guaranteed to have a timestamp no less than than anything in its causal history.

\subsection{Update Functions}
\label{subsec:clocks-spanning}

Given a logical clock, our goal is to derive a global assignment of timestamps to events for any CSD.
Following the pattern in \Cref{sec:interpreting-csds}, we first restrict our attention to an assignment of timestamps to the \emph{bounding sites} of an \(\mathrm{Act}\)-labeled diagram \(X : \Gamma_1 \rightrightarrows^\mathrm{Act} \Gamma_2\).

Intuitively, we will want to interpret every \(\langle \mathbf{tick} , a \rangle\) as an \(\mathbf{increment}_a\) operation, and every \(\langle \mathbf{join} , \bullet \rangle\) as a \(\mathbf{merge}\) over the input timestamps.
An \(\mathrm{Act}\)-labeled CSD is then an expression arranging any number of clock operations on timestamps into a one-shot, compound operation over an entire configuration of clocks.
In other words, every \(\mathrm{Act}\)-labeled CSD yields a function mapping an assignment of timestamps on its input sites to an assignment of timestamps on its output sites.

\begin{definition}[Valuations]
  \label{def:valuations}
  The type of \textit{valuations on \(\Gamma\)}, written \(\mathrm{Valuation}(\Gamma)\), is the type of functions \(\nu : \mathrm{Site}(\Gamma) \rightarrow \mathrm{Time}\) assigning a timestamp to each site in \(\Gamma\).
\end{definition}

\begin{definition}[Update functions]
  \label{def:interp-functions}
  For every logical clock, the interpretation \(\llbracket - \rrbracket\) of \(\mathrm{Act}\)-labeled CSDs \(X : \Gamma_1 \rightrightarrows^\mathrm{Act} \Gamma_2\) into \emph{update functions} of type \(\mathrm{Valuation}(\Gamma_1) \rightarrow \mathrm{Valuation}(\Gamma_2)\) is defined as:
  \begin{alignat*}{5}
    &   \llbracket \mathbf{tick} , \bullet \mapsto a \rrbracket
    &&= \lambda \nu.\ \lambda {-}.\ \mathbf{increment}_a(\nu(\bullet))
  \\
    &   \llbracket \mathbf{fork} , {-} \rrbracket
    &&= \lambda \nu.\ \lambda {-}.\ \nu(\bullet)
  \\
    &   \llbracket \mathbf{join} , {-} \rrbracket
    &&= \lambda \nu.\ \lambda {-}.\ \nu(\mathbf{inj}_\ell(\bullet)) \sqcup \nu(\mathbf{inj}_r(\bullet))
  \\
    &   \llbracket \mathbf{perm}\ \sigma , {-} \rrbracket
    &&= \lambda \nu.\ \nu \circ \sigma^{-1}
  &\quad
    &   \llbracket \mathbf{id} , {-} \rrbracket
    &&= \lambda \nu.\ \nu
  \\
    &   \llbracket x \parallel y , f_x + f_y \rrbracket
    &&= \llbracket x , f_x \rrbracket + \llbracket y , f_y \rrbracket
  &\quad
    &   \llbracket x \mathbin{;} y , f_x + f_y \rrbracket
    &&= \llbracket y , f_y \rrbracket \circ \llbracket x , f_x \rrbracket
  \end{alignat*}
  When the diagram \(X\) is understood, we will write \(\mathcal{C}_\nu[s]\) to mean \(\llbracket X \rrbracket(\nu)(s)\).
\end{definition}

Because a \(\mathbf{tick}\) transforms a valuation on one site into a valuation on one site, it serves as a very thin wrapper around \(\mathbf{increment}_a\).
The new valuation can ignore its argument, because there is only one input to a \(\mathbf{tick}\).
Likewise, \(\mathbf{fork}\) ignores its argument because both outputs receive their timestamp from the same input site, and \(\mathbf{join}\) merges both input sites onto the single output site.

In contrast, the \(\mathbf{perm}\) constructor doesn't manipulate any timestamps directly.
Instead, any given site is translated by the permutation \(\sigma\) into an index on the input valuation: the requested timestamp is just one of those in the input.
The \(\mathbf{id}\) constructor behaves similarly.

Finally, sequential and concurrent composition each combine the evaluation functions from each subdiagram.
Sequential composition is given by the usual composition of functions (\(\circ\)); and concurrent composition is given by the usual pairing of two functions over a sum type (\(+\)).
We abuse pattern-matching notation somewhat by writing \(f_x + f_y\) on the left-hand side, where we would otherwise write simply \(f\) and compose its uses with the appropriate injection.

\subsection{Clock Functions}
\label{subsec:clocks-interior}

The interpretation of \Cref{def:interp-functions} only tells us what timestamps a system terminates on, not the timestamps along the way.
To obtain the latter, we must extend our function \(\mathcal{C}_\nu\) to accept any event (\Cref{def:events}), not just output sites.
That is, we want a function \(\mathcal{C} : \mathrm{Valuation}(\Gamma_1) \rightarrow (\mathrm{Event}(X) \rightarrow \mathrm{Time})\), computing an assignment of timestamps to all events given an initial assignment of timestamps.

Following \Cref{subsec:interior-paths}, we will select a subdiagram with the event of interest on its boundary.
The timestamp at an event is then one of the timestamps on which that subdiagram terminates.

\begin{definition}[The subdiagram before a cut]
  \label{def:subdiagrams-before}
  The \emph{subdiagram before a cut} \(t\), denoted \(\mathit{before}(t)\), is the CSD consisting of the global steps appearing strictly before the cut \(t\) in a diagram \(X\).
\end{definition}

\begin{definition}[Clock function]
  \label{def:clock-function}
  For every choice of logical clock and \(\mathrm{Act}\)-labeled diagram \(X\), the \emph{clock function} \(\mathcal{C}\) of type \(\mathrm{Valuation}(\Gamma_1) \rightarrow (\mathrm{Event}(X) \rightarrow \mathrm{Time})\) is given by
  \begin{alignat*}{2}
    &   \mathcal{C}_\nu[(t, s)]
    &&= \llbracket \mathit{before}(t) \rrbracket(\nu)(s).
  \end{alignat*}
\end{definition}

We consistently pun \(\mathcal{C}_\nu\) to mean either the update function (\Cref{def:interp-functions}) or the clock function depending on whether its argument is a site or an event.

\begin{figure}
  \begin{center}
    \includegraphics[width=\columnwidth]{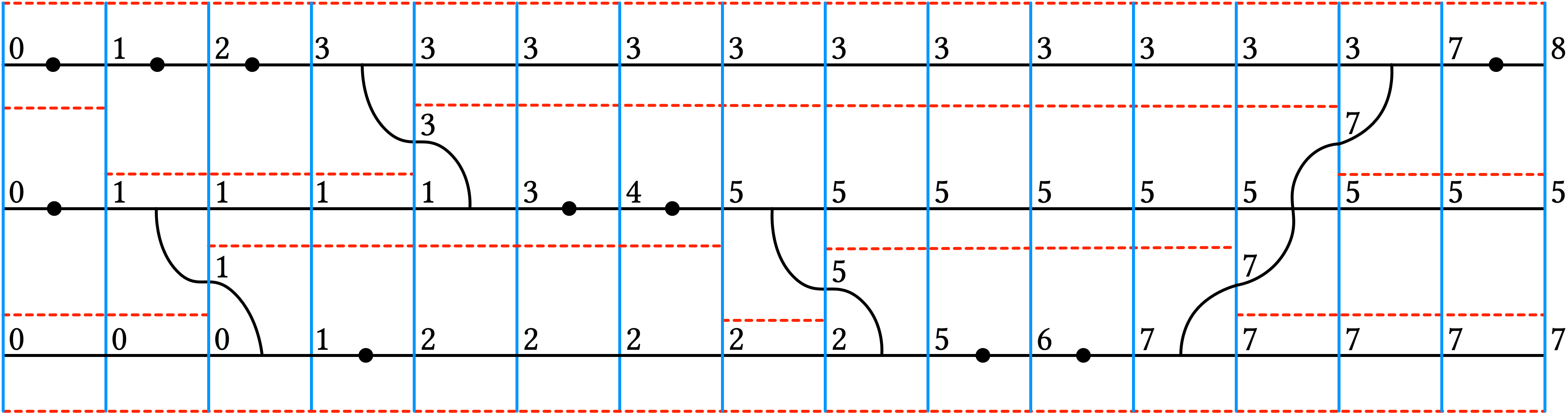}
  \end{center}
  \caption{The execution from \Cref{fig:clock-examples} as a CSD, with Lamport timestamps assigned to events.}
  \label{fig:labeled-csd-example}
\end{figure}

\Cref{fig:labeled-csd-example} depicts the execution from \Cref{fig:clock-examples} as a CSD, with timestamps assigned to events according to the Lamport clock, given a starting valuation of zeroes and using the interpretation in \Cref{def:interp-functions}.
As discussed in \Cref{subsec:interior-paths}, we can associate timestamps to \emph{actions} rather than events just by selecting one of the neighboring events for each action to represent it.
In this case, convention suggests adopting the timestamp of the event immediately following each action.

\section{Relating Causal Paths to Clocks}
\label{sec:interp-proofs}

In \Cref{sec:interp-causal-paths}, we introduced an interpretation into paths \(e_1 \rightsquigarrow e_2\), giving a proof-relevant causal order on events; and in \Cref{sec:interp-clock-programs}, we introduced a family of interpretations into clock functions \(\mathcal{C}_\nu[-]\), giving an assignment of timestamps to events.
In this section, we will relate these two interpretation via a third, ultimately yielding a proof of the clock condition: if \(e_1 \rightsquigarrow e_2\), then \(\mathcal{C}_\nu[e_1] \le \mathcal{C}_\nu[e_2]\).
Following the recipe in \Cref{sec:interpreting-csds}, we will again begin with a \emph{spanning} proof relating paths and timestamps on the bounding sites, then extend to an \emph{interior} proof relating paths and timestamps on all events.

\subsection{Inflationarity of Update Functions}
\label{subsec:proofs-spanning}

The clock condition relates any two events in a diagram: if \(e_1 \rightsquigarrow e_2\), then \(\mathcal{C}_\nu[e_1] \le \mathcal{C}_\nu[e_2]\).
If we restrict our attention to sites \(s_1, s_2\) at the start and end of the diagram, respectively, then \(\mathcal{C}_\nu[e_1]\) reduces to simply \(\nu(s_1)\), because the diagram before an initial site is the empty diagram \(\mathbf{id}\).
This leads us to the following statement:

\begin{theorem}[The update function is inflationary]
  \label{thm:update-inflationary}
  Fix a choice of logical clock, and let \(X\) be an \(\mathrm{Act}\)-labeled CSD \(\Gamma_1 \rightrightarrows^\mathrm{Act} \Gamma_2\) with an initial valuation \(\nu : \mathrm{Valuation}(\Gamma_1)\).
  Then the clock's update function \(\mathcal{C}\) is inflationary on causally related sites:
  \[
    \forall (s_1 : \mathrm{Site}(\Gamma_1)) (s_2 : \mathrm{Site}(\Gamma_2).\ (s_1 \rightsquigarrow s_2) \rightarrow (\nu(s_1) \le \mathcal{C}_\nu[s_2]).
  \]
\end{theorem}

This property is an analogue of the inflationary property satisfied by the clock operations of \Cref{def:logical-clocks}: if an output \emph{can be} influenced by an input, then the output \emph{must be} bounded below by the input.
In some ways, it would be surprising if \Cref{thm:update-inflationary} didn't hold of \(\mathcal{C}\), as it is built entirely from inflationary clock operations.
Our proof will be built in kind, composing proofs over atomic steps to yield proofs for entire diagrams.
We sketch the proof at a high level here; the details are available in our Agda development.

\begin{itemize}
  \item
    The proof for a \(\mathbf{tick}\) step uses the fact that the clock's \(\mathbf{increment}\) operator is inflationary: \(t \le \textbf{increment}_a(t)\) for every action \(a\) and timestamp \(t\).
    This is true by construction for any clock implementing \Cref{def:logical-clocks}.

  \item
    The proof for a \(\mathbf{join}\) step uses the fact that the clock's \(\sqcup\) operator is inflationary on both arguments: both \(t_1 \le (t_1 \sqcup t_2)\) and \(t_2 \le (t_1 \sqcup t_2)\) for every pair of timestamps \(t_1\), \(t_2\).
    Again, this is definitionally true.

  \item
    The proof for a \(\mathbf{fork}\) step uses the fact that the clock's ordering relation is reflexive: we simply copy the input timestamp onto both outputs, so the actual values are unchanged.
    Indeed, this is true of \(\mathbf{perm}\) and \(\mathbf{id}\), too: all outputs are precisely the same as the (unique) inputs they are causally related to.

  \item
    The proof for a sequential composition (\(\mathbin{;}\)) uses the fact that the clock's ordering relation is transitive.
    If we have a path through an intermediate site, where the time at the intermediate site is bounded below at the input and bounded above at the output, we must use transitivity to obtain a direct relationship between the input and output.

  \item
    The proof for a concurrent composition requires no information about the clock; however, the \emph{proof-relevance} of our causal relation plays an essential role.
    We know that \(s_1\) and \(s_2\) are causally ordered because we were given a \emph{specific} path witnessing the fact; and any given path through a concurrent composition is a path wholly through one concurrent half of the diagram or the other.
    Thus, we can simply dispatch to whichever sub-proof applies to the path at hand.
\end{itemize}

Somewhat surprisingly, nowhere do we require antisymmetry: even though partial orders are traditionally used in logical clocks, \emph{preorders} are enough.
This proof also holds for \emph{every} CSD, even those not reflecting a well-behaved system.
All we require is that updates are inflationary --- the clock condition is not actually sensitive to \emph{what} those updates are, or \emph{who} performs them.
This reveals a clean separation between clocks as ADTs and the protocols they are employed in; the clock condition is solely concerned with the ADT itself.

\subsection{Monotonicity of Clock Functions}
\label{subsec:proofs-interior}
\label{forward:generic-causality-proof}

Just as in \Cref{subsec:interior-paths,subsec:clocks-interior}, we need to be a little creative to leverage \Cref{thm:update-inflationary} into a proof of the full clock condition.
The key insight is that, if we have a path of type \(e_1 \rightsquigarrow e_2\) and an initial valuation \(\nu\), we can run the clock's update function on the subdiagram \emph{before} \(e_1\).
The resulting valuation is an initial valuation for the subdiagram \emph{between} \(e_1\) and \(e_2\), on which we can apply inflationarity.
Once more, we leave the finer details to our Agda implementation.

\begin{theorem}[The clock function is monotonic]
  \label{thm:clock-monotonic}
  Fix a choice of logical clock, and let \(X\) be an \(\mathrm{Act}\)-labeled CSD \(\Gamma_1 \rightrightarrows^\mathrm{Act} \Gamma_2\) with an initial valuation \(\nu : \mathrm{Valuation}(\Gamma_1)\).
  Then the clock function \(\mathcal{C}\) is monotonic on causally related events:
  \[
    \forall (e_1\ e_2 : \mathrm{Event}(X)).\ (e_1 \rightsquigarrow e_2) \rightarrow (\mathcal{C}_\nu[e_1] \le \mathcal{C}_\nu[e_2]).
  \]
\end{theorem}

\Cref{thm:clock-monotonic} tells us that every logical clock implementing the clock ADT of \Cref{def:logical-clocks} must necessarily satisfy the clock condition.
Notably, this theorem applies to \emph{all} CSDs, even those that may be produced by clock implementations that may be incorrect in certain ways (e.g. a process incrementing the wrong component of a timestamp).
That is, the clock condition holds by virtue of its interface as an abstract data type, not merely in the context of a well-behaved client program.
In \Cref{sec:verified-clocks}, we will actually instantiate these results on several clocks from the literature.

\section{Verified Logical Clocks}
\label{sec:verified-clocks}

In \Cref{sec:interp-causal-paths,sec:interp-clock-programs,sec:interp-proofs}, we developed a framework for reasoning about causal relationships and logical clocks, culminating in a generic proof of the clock condition for implementations of the standard clock abstract data type.
In this section we apply our results to several well-known clocks: \citeauthor{lamport-clocks}'s scalar clock~[\citeyear{lamport-clocks}], \citeauthor{mattern-vector-time}'s vector clock~[\citeyear{mattern-vector-time}], \citeauthor{rst-matrix-clock}'s matrix clock~[\citeyear{rst-matrix-clock}], and \citeauthor{wuu-bernstein-matrix-clocks}'s matrix clock [\citeyear{wuu-bernstein-matrix-clocks}].
Implementations of these clocks are included in our Agda development, each with an instantiation of our generic proof of the clock condition.

Although there is only one ``scalar'' clock and ``vector'' clock in common use, there are two distinct ``matrix'' clocks with two-dimensional timestamps.
The clock of \citeauthor{rst-matrix-clock}, like the others we discuss, merges timestamps strictly pointwise; in contrast, the clock of \citeauthor{wuu-bernstein-matrix-clocks} additionally merges a row at one index into a row at another, yielding a \emph{noncommutative} merge operator.
To avoid confusion, we will refer to the former as \emph{the RST clock}, and the latter as \emph{the Wuu-Bernstein clock}.
We will have more to say about the characteristics of the Wuu-Bernstein clock in \Cref{subsec:verifying-tensor-clocks}; for now, we restrict our attention to the scalar, vector, and RST clocks.

\subsection{Classifier Clocks}
\label{subsec:verifying-classifier-clocks}

The scalar, vector, and RST clocks all follow a similar template: we \emph{classify} actions by some domain-specific criterion, then maintain a count of observed actions for every class.
\begin{itemize}
  \item The scalar clock classifies all actions into one single, universal class. Its timestamp consists of a single natural number, assessing a lower bound on the total number of actions that have occurred prior.

  \item The vector clock classifies actions based on who performed them, i.e. by \emph{actor}. Its timestamp consists of a vector of natural numbers --- or, equivalently, a function assigning a natural to every actor.

  \item The RST clock classifies actions based on \emph{subject} and \emph{object}: that is, every action is performed by some subject against some object.
  For \citet{rst-matrix-clock}, these actions are the submission of messages, where every message has both a sender (the subject) and a recipient (the object).
  The RST clock's timestamp is thus a table counting messages sent between any two actors --- or, equivalently, a function assigning a natural to every pair of actors.
\end{itemize}
Surprisingly, these clocks turn out to be structurally identical, differing only in their indexing classes \(I\).
In all cases, timestamps are maps \(I \rightarrow \mathbb{N}\) ordered pointwise; the \(\mathbf{increment}\) operation increments the value for a chosen class \(i \in I\) by one; and the merge of two timestamps is their pointwise maximum.
From elementary properties of natural numbers, this pointwise order is a preorder, and both operations are inflationary.
Thus, we model all three clocks with one implementation, which we call a \textbf{classifier clock}, parametric in a classification function giving each action its class.

By instantiating \Cref{def:clock-function} and \Cref{thm:clock-monotonic} on the classifier clock, we obtain a global assignment of timestamps for every CSD, together with a proof that this assignment is monotone (i.e., the clock condition).
\begin{corollary}[Clock condition for classifier clocks]
  Every classifier clock whose operations \(\mathbf{increment}_a\) and \(\sqcup\) are inflationary satisfies the clock condition.
\end{corollary}
When specialized to sender-recipient classes (that is, indices \(\mathrm{Pid} \times \mathrm{Pid}\)), this yields the first mechanized proof (to our knowledge) of the clock condition for the RST clock.

\subsection{Tensor Clocks}
\label{subsec:verifying-tensor-clocks}

The \citet{wuu-bernstein-matrix-clocks} clock differs from the others in that it merges a row at the sender's index into a row at the recipient's, in addition to the usual pointwise merge.
This merge operation is noncommutative, since it depends on which timestamp is considered the sender's, and which is considered the recipient's.

\citet{kshemkalyani-tensor-clocks} constructs a whole \emph{tensor clock hierarchy} of clocks with noncommutative merge, where a general index \((c, o_1, o_2, \dots)\) models information of the form ``\(o_1\) knows that \(o_2\) knows that \(\dots\) \(c\) occurred at least \(N\) many times.''
These clocks model a kind of transitive knowledge: if one agent observes some population of actions, and they send a message to another agent, then the recipient transitively observes that same population of actions.
The \citeauthor{wuu-bernstein-matrix-clocks} clock falls out as a special case of the tensor clock hierarchy
\footnote{The vector clock also appears as a member of the tensor clock hierarchy, though it exists as something of a base case --- unlike higher tensor clocks, its merge is commutative.}%
, and it --- along with all other tensor clocks --- satisfies the clock condition via \cref{thm:clock-monotonic} despite noncommutative merge.
\begin{corollary}[Clock condition for tensor clocks]
  Every tensor clock whose operations \(\mathbf{increment}_a\) and \(\sqcup\) are inflationary satisfies the clock condition.
\end{corollary}

We have implemented and verified the clock condition for the Wuu-Bernstein clock in our framework.
However, the noncommutative merge operation poses some theoretical problems for the model of interpretation we developed in \Cref{sec:interp-clock-programs}, which interprets the \(\mathbf{join}\) atomic step into the clock's merge operator.
We want to treat \(\mathbf{join}\) as commutative (up to isomorphism), as with the products of sets or types.
Therefore, an interpretation via \Cref{def:interp-functions} of \(\mathbf{join}\) into a noncommutative merge operator would take equivalent CSDs to non-equivalent update functions.
That said, since all such update functions are increasing, our proof of the clock condition in \Cref{thm:clock-monotonic} still holds --- there is no pair of equivalent CSDs for which the clock condition holds on one but not the other.
Nonetheless, we hope to construct a more adequate interpretation that accounts for the full tensor clock hierarchy in the future.

\section{Related Work}
\label{sec:related-work}

\paragraph{MSCs and their semantics.}

Message sequence charts (MSCs) \citep{msc-standard} are a diagrammatic language for representations of message-passing computations, widely used by practitioners and researchers (e.g.,~\citet{lohrey-mscs,alur-mscs,bollig-mscs,di-giusto-mscs}, as a small sampling).
MSCs are closely related to Lamport diagrams, being defined in terms of straight-line processes and messages crossing between them.
There have been various efforts to formalize MSCs or MSC-like diagrammatic languages, including the MSC standard itself~\citep{msc-standard} and others~\citep{schaltz-eets}, and investigations of the semantics of MSCs~\citep{ladkin-msc-semantics,broy-msc-semantics,alur-msc-analyzer,mauw-reniers-msc-semantics,gehrke-msc-semantics}.
However, we are not aware of any formalizations of MSCs that define them inductively, as we have defined CSDs.

\citet{alur-msc-analyzer} note that MSCs admit ``a variety of semantic interpretations'', seemingly similar in spirit to our interpretations of CSDs.
However, \citeauthor{alur-msc-analyzer}'s interpretations yield refinements of causal order -- for example, they note that the meaning of a given MSC may depend on the choice of network model and fault model (e.g., whether message loss or reordering are possible).
While we give an interpretation of CSDs into a causal order, our range of possible semantic domains is greater: we also give interpretations into computable functions and into proofs.

\paragraph{Mechanized reasoning about clocks and causality in concurrent systems.}

In distributed systems, the notion of causal ordering arises in a myriad of settings, including causally consistent data stores~\citep{ahamad-causal-memory, lloyd-cops},
distributed snapshot protocols~\citep{mattern-vector-time,acharya-causal-snapshots,alagar-causal-snapshots}, causal message delivery protocols~\citep{birman-virtual-synchrony, schiper-causal-ordering, birman-reliable, birman-lightweight-cbcast},
and conflict-free replicated data types (CRDTs)~\citep{shapiro-crdts}.
In shared-memory systems, the need to reason about causality arises in the setting of data race detection for multithreaded programs~\citep{pozniansky-djit,flanagan-fasttrack}.
It is typical for such applications to use logical clocks of one kind or another to reify causal information.

There are several mechanically verified implementations of distributed algorithms that use logical clocks~\citep{lesani-chapar,gondelman-distributed-causal-memory, nieto-verified-crdts,redmond-verified-causal-broadcast}.
These proof developments focus on verifying properties of those higher-level algorithms (such as causal consistency of replicated databases~\citep{lesani-chapar, gondelman-distributed-causal-memory}, convergence of CRDTs~\citep{nieto-verified-crdts}, or safety of causal message broadcast~\citep{nieto-verified-crdts,redmond-verified-causal-broadcast}), and they (implicitly or explicitly) take the clock condition as an axiom.
Our mechanized proof of the clock condition is \emph{generic} for any clock that can be realized by a system of runtime replicas --- in other words, a clock defined in terms of standard ``increment'' and ``merge'' functions.

The only other work that we are aware of on mechanized verification of the clock condition itself is by \citet{mansky-verified-clocks}, whose work focuses on the verification of dynamic race detection algorithms.
As part of their larger proof development, \citeauthor{mansky-verified-clocks} proved in Coq that vector clocks precisely characterize the causal order.
That is, they proved not only the clock condition for vector clocks, as we do here, but also the \emph{inverse} clock condition: if \(e_i\)'s timestamp is less than \(e_j\)'s timestamp, then \(e_i\) causally precedes \(e_j\).
Unlike the (forward) clock condition, the inverse clock condition depends on the particular protocol governing use of the clock: a process must not increment an index owned by another process.
While we verified the clock condition for a whole class of clocks, we do not yet attack protocol-dependent properties like the inverse clock condition, though we hope to do so eventually.

\paragraph{Formal models for reasoning about protocols.}

\citet{talupur-tuttle-message-flows} introduce \emph{message flows} as a methodology for formal reasoning about distributed protocols.
They observe that execution diagrams, such as Lamport diagrams and message sequence charts, need not be limited to informal reasoning on whiteboards, but can be taken seriously as mathematical objects.
That has been our intention with CSDs, as well.
More recently, \citet{mora-message-chains} present a verification methodology based on \emph{message chains}, which reveal causal structure in executions of distributed systems, allowing protocol designers to reason about system behavior at a high level of abstraction.
Message flows and message chains could likely be modeled as inductive paths in our formalism.
CSDs provide a rich model of executions on which these verification techniques could potentially be built more easily.

The Logic of Events (LoE) of \citet{bickford-logic-events} builds on a Lamport-style model of executions in support of analysis and synthesis of distributed programs given an event-based specification.
The EventML system~\cite{rahli-eventml} builds on LoE by (among other contributions) incorporating a process model for implementations satisfying LoE formulae, as well as a high-level language (the titular ``EventML'') that lowers to both LoE and their process model.
A key capability of EventML is its generation of inductive properties characterizing the behavior of an EventML program specification, supporting local reasoning about operations in terms of causally-available inputs.
In comparison, CSDs provide an inductive model of single-run executions that each depict one possible behavior.
While CSDs also emphasize induction, our induction is structural over the CSD itself, while EventML / LoE formulae appear to be inductive in the sense of a property over a well-founded set: a property that holds at one time will hold at all prior times.
In principle, CSDs might be given interpretations \emph{into} a domain of inductive properties; but EventML's inductive properties describe whole programs (i.e. a whole class of runs), where induction over CSDs proceeds over individual runs.
Nonetheless, we hope to investigate connections with program- and protocol-level properties in the future, at which point a connection with LoE and EventML might become more apparent.

\paragraph{Separation logics.}

Separation logics~\citep{reynolds-separation-logic} are program logics for reasoning about the correct use of resources --- concrete resources such as memory, but, excitingly, also \emph{logical resources} such as permissions and execution history.
\emph{Concurrent} separation logics~\citep{ohearn-csl} enable such reasoning about concurrent programs.
The literature on separation logics and concurrent separation logics is too vast to summarize here, although \citet{ohearn-cacm-sl} offers an accessible introduction and \citet{brookes-ohearn-csl} give an overview of important developments.
CSDs are heavily inspired by concurrent separation logic, but we have not yet pursued a program logic based on CSDs.
\citet{wickerson-ribbon-proofs}'s \emph{ribbon proofs}, a diagrammatic proof system based on separation logic, could be an inspiration for future work in this direction.

Separation logic has been used in the service of reasoning about causality.
\citet{gondelman-distributed-causal-memory} and \citet{nieto-verified-crdts} both use the Aneris concurrent separation logic framework~\citep{krogh-jespersen-aneris}, itself built on the Iris~\citep{jung-iris} framework, to verify the correctness of distributed systems in which causality is a central concern.
However, the Aneris framework does not offer any particular support for reasoning about causality.
In fact, we are not aware of program logics or verification frameworks that are specifically intended for reasoning about causality, which is perhaps surprising, considering the importance of causality in concurrent systems.
Rather than reasoning about causal relationships as logical resources, as one would do when using Iris or Aneris, causality in a CSD-based proof system would manifest in the structure of the proof itself.

\paragraph{String diagrams.}

Our CSDs are inspired by the \emph{string diagrams} employed in category theory, which formally describe compositions of morphisms in a monoidal category (i.e., with a concurrent composition operator) using a graphical syntax.
The standard reference for string diagrams is \citet{joyal-street-geometry}, though \citet{piedeleu-intro-string-diagrams} give an accessible introduction for computer scientists.
We hope to establish firmer connections between CSDs and string diagrams in future work, e.g., by proving that CSDs form a (symmetric) monoidal category.
Moreover, recent work by \citet{nester-concurrent-histories} has described execution traces in concurrent systems using string diagrams in which data can be transferred between tiles both in time (in the forward direction) and in space (in the sideways direction).
This contrasts with our CSDs, in which data is only transferred in time.
\citeauthor{nester-concurrent-histories} leverages \emph{double categories} to formalize these two-dimensional interfaces.
It would be interesting to investigate what a treatment of causality might look like in such a setting.

String diagrams have been applied to many domains beyond distributed systems.
In quantum computing, the ZX-calculus~\citep{coecke-zx-calculus} is a graphical formalism for the description of quantum circuits, much like Lamport diagrams are a graphical formalism for distributed traces.
The VyZX project~\citep{lehmann-vyzx,lehmann-vyzx-2023} has mechanized the ZX-calculus as an inductive data type, motivated (as we were) by the desire to do inductive reasoning in a mechanized setting, and to ground their diagrams against a variety of semantics.
VyZX emphasizes the application of rewrite rules to prove equivalence of diagrams, whereas we have emphasized the connectivity between events in a diagram.
Similarly, the tape diagrams of \citet{bonchi-tape-diagrams} give a graphical syntax to set-theoretic relations.
A tape diagram is a two-layer presentation of relations, with disjunction on one layer and conjunction on another.
The two-layer structure of CSDs, with global actions over global state decomposing into local actions over local state, is reminiscent of \citeauthor{bonchi-tape-diagrams}'s approach.
We would like to explore the connections among these ideas in future work.

\section{Conclusion}
\label{sec:discussion}

Causality is of central importance in concurrent systems, including both shared-state and message-passing systems.
In this paper, we presented causal separation diagrams (CSDs), a new formal model of concurrent executions that is inductively defined and enjoys a diagrammatic syntax reminiscent of Lamport diagrams.
The inductive nature of CSDs makes them amenable to \emph{mechanized reasoning} and \emph{interpretation}.

As a case study, we used CSDs to reason about logical clocks, ubiquitous mechanisms for reifying causal information in concurrent systems.
By interpreting CSDs into a variety of semantic domains, we built up a generic proof of Lamport's clock condition that holds for any realizable logical clock, including the Wuu-Bernstein clock and the RST clock, neither of which were mechanically verified previously.
A proof-relevant analogue of Lamport's \happensbefore{} relation, witnessing concrete causal paths in an execution, plays an essential role in these proofs.
Our framework and results are available as an Agda development.

While logical clocks were a focus of this paper, we see CSDs (and interpretations of CSDs) as a valuable reasoning tool beyond their application to logical clocks.
In future work, we hope to flesh out the connection between CSDs and symmetric monoidal categories in more detail, including notions of equivalence and refinement for CSDs, which will hopefully yield well-behavedness conditions for interpretations.

\section*{Acknowledgements}
\label{sec:acks}

We thank the anonymous reviewers of POPL '24 and OOPSLA '24 for their feedback, without which our presentation would have been worse off. Discussions with Ryan Doenges, Ilya Sergey, and the VyZX team aided our understanding of related work.
Gan Shen and Simon Guo were an early source of inspiration for our analysis of logical clocks.

This material is based upon work supported by the National Science Foundation under Grant
No. CCF-2145367. Any opinions, findings, and conclusions or recommendations expressed in this
material are those of the author(s) and do not necessarily reflect the views of the National Science
Foundation.

\bibliographystyle{ACM-Reference-Format}
\bibliography{references}


\begin{thebibliography}{60}


\ifx \showCODEN    \undefined \def \showCODEN     #1{\unskip}     \fi
\ifx \showDOI      \undefined \def \showDOI       #1{#1}\fi
\ifx \showISBNx    \undefined \def \showISBNx     #1{\unskip}     \fi
\ifx \showISBNxiii \undefined \def \showISBNxiii  #1{\unskip}     \fi
\ifx \showISSN     \undefined \def \showISSN      #1{\unskip}     \fi
\ifx \showLCCN     \undefined \def \showLCCN      #1{\unskip}     \fi
\ifx \shownote     \undefined \def \shownote      #1{#1}          \fi
\ifx \showarticletitle \undefined \def \showarticletitle #1{#1}   \fi
\ifx \showURL      \undefined \def \showURL       {\relax}        \fi
\providecommand\bibfield[2]{#2}
\providecommand\bibinfo[2]{#2}
\providecommand\natexlab[1]{#1}
\providecommand\showeprint[2][]{arXiv:#2}

\bibitem[Acharya and Badrinath(1992)]%
        {acharya-causal-snapshots}
\bibfield{author}{\bibinfo{person}{Arup Acharya} {and} \bibinfo{person}{B.R.
  Badrinath}.} \bibinfo{year}{1992}\natexlab{}.
\newblock \showarticletitle{Recording distributed snapshots based on causal
  order of message delivery}.
\newblock \bibinfo{journal}{\emph{Inform. Process. Lett.}}
  \bibinfo{volume}{44}, \bibinfo{number}{6} (\bibinfo{year}{1992}),
  \bibinfo{pages}{317--321}.
\newblock
\showISSN{0020-0190}
\urldef\tempurl%
\url{https://doi.org/10.1016/0020-0190(92)90107-7}
\showDOI{\tempurl}


\bibitem[Ahamad et~al\mbox{.}(1995)]%
        {ahamad-causal-memory}
\bibfield{author}{\bibinfo{person}{Mustaque Ahamad}, \bibinfo{person}{Gil
  Neiger}, \bibinfo{person}{James~E. Burns}, \bibinfo{person}{Prince Kohli},
  {and} \bibinfo{person}{Phillip~W. Hutto}.} \bibinfo{year}{1995}\natexlab{}.
\newblock \showarticletitle{Causal memory: definitions, implementation, and
  programming}.
\newblock \bibinfo{journal}{\emph{Distributed Computing}} \bibinfo{volume}{9},
  \bibinfo{number}{1} (\bibinfo{year}{1995}), \bibinfo{pages}{37--49}.
\newblock
\showISBNx{1432-0452}
\urldef\tempurl%
\url{https://doi.org/10.1007/BF01784241}
\showDOI{\tempurl}


\bibitem[Alagar and Venkatesan(1994)]%
        {alagar-causal-snapshots}
\bibfield{author}{\bibinfo{person}{Sridhar Alagar} {and} \bibinfo{person}{S.
  Venkatesan}.} \bibinfo{year}{1994}\natexlab{}.
\newblock \showarticletitle{An optimal algorithm for distributed snapshots with
  causal message ordering}.
\newblock \bibinfo{journal}{\emph{Inform. Process. Lett.}}
  \bibinfo{volume}{50}, \bibinfo{number}{6} (\bibinfo{year}{1994}),
  \bibinfo{pages}{311--316}.
\newblock
\showISSN{0020-0190}
\urldef\tempurl%
\url{https://doi.org/10.1016/0020-0190(94)00055-7}
\showDOI{\tempurl}


\bibitem[Altenkirch and Morris(2009)]%
        {altenkirch-containers}
\bibfield{author}{\bibinfo{person}{Thorsten Altenkirch} {and}
  \bibinfo{person}{Peter Morris}.} \bibinfo{year}{2009}\natexlab{}.
\newblock \showarticletitle{Indexed Containers}. In
  \bibinfo{booktitle}{\emph{2009 24th Annual IEEE Symposium on Logic In
  Computer Science}}. \bibinfo{pages}{277--285}.
\newblock
\urldef\tempurl%
\url{https://doi.org/10.1109/lics.2009.33}
\showDOI{\tempurl}


\bibitem[Alur et~al\mbox{.}(2000)]%
        {alur-mscs}
\bibfield{author}{\bibinfo{person}{Rajeev Alur}, \bibinfo{person}{Kousha
  Etessami}, {and} \bibinfo{person}{Mihalis Yannakakis}.}
  \bibinfo{year}{2000}\natexlab{}.
\newblock \showarticletitle{Inference of Message Sequence Charts}. In
  \bibinfo{booktitle}{\emph{Proceedings of the 22nd International Conference on
  Software Engineering}} (Limerick, Ireland) \emph{(\bibinfo{series}{ICSE
  '00})}. \bibinfo{publisher}{Association for Computing Machinery},
  \bibinfo{address}{New York, NY, USA}, \bibinfo{pages}{304–313}.
\newblock
\showISBNx{1581132069}
\urldef\tempurl%
\url{https://doi.org/10.1145/337180.337215}
\showDOI{\tempurl}


\bibitem[Alur et~al\mbox{.}(1996)]%
        {alur-msc-analyzer}
\bibfield{author}{\bibinfo{person}{Rajeev Alur}, \bibinfo{person}{Gerard~J.
  Holzmann}, {and} \bibinfo{person}{Doron Peled}.}
  \bibinfo{year}{1996}\natexlab{}.
\newblock \showarticletitle{An analyzer for message sequence charts}. In
  \bibinfo{booktitle}{\emph{Tools and Algorithms for the Construction and
  Analysis of Systems}}, \bibfield{editor}{\bibinfo{person}{Tiziana Margaria}
  {and} \bibinfo{person}{Bernhard Steffen}} (Eds.).
  \bibinfo{publisher}{Springer Berlin Heidelberg}, \bibinfo{address}{Berlin,
  Heidelberg}, \bibinfo{pages}{35--48}.
\newblock
\showISBNx{978-3-540-49874-2}


\bibitem[Bickford(2009)]%
        {bickford-logic-events}
\bibfield{author}{\bibinfo{person}{Mark Bickford}.}
  \bibinfo{year}{2009}\natexlab{}.
\newblock \showarticletitle{Component Specification Using Event Classes}. In
  \bibinfo{booktitle}{\emph{Component-Based Software Engineering}},
  \bibfield{editor}{\bibinfo{person}{Grace~A. Lewis}, \bibinfo{person}{Iman
  Poernomo}, {and} \bibinfo{person}{Christine Hofmeister}} (Eds.).
  \bibinfo{publisher}{Springer Berlin Heidelberg}, \bibinfo{address}{Berlin,
  Heidelberg}, \bibinfo{pages}{140--155}.
\newblock
\showISBNx{978-3-642-02414-6}


\bibitem[Birman and Joseph(1987a)]%
        {birman-virtual-synchrony}
\bibfield{author}{\bibinfo{person}{K. Birman} {and} \bibinfo{person}{T.
  Joseph}.} \bibinfo{year}{1987}\natexlab{a}.
\newblock \showarticletitle{Exploiting Virtual Synchrony in Distributed
  Systems}.
\newblock \bibinfo{journal}{\emph{SIGOPS Oper. Syst. Rev.}}
  \bibinfo{volume}{21}, \bibinfo{number}{5} (\bibinfo{date}{Nov.}
  \bibinfo{year}{1987}), \bibinfo{pages}{123–138}.
\newblock
\showISSN{0163-5980}
\urldef\tempurl%
\url{https://doi.org/10.1145/37499.37515}
\showDOI{\tempurl}


\bibitem[Birman et~al\mbox{.}(1991)]%
        {birman-lightweight-cbcast}
\bibfield{author}{\bibinfo{person}{Kenneth Birman}, \bibinfo{person}{Andr\'{e}
  Schiper}, {and} \bibinfo{person}{Pat Stephenson}.}
  \bibinfo{year}{1991}\natexlab{}.
\newblock \showarticletitle{Lightweight Causal and Atomic Group Multicast}.
\newblock \bibinfo{journal}{\emph{ACM Trans. Comput. Syst.}}
  \bibinfo{volume}{9}, \bibinfo{number}{3} (\bibinfo{date}{Aug.}
  \bibinfo{year}{1991}), \bibinfo{pages}{272–314}.
\newblock
\showISSN{0734-2071}
\urldef\tempurl%
\url{https://doi.org/10.1145/128738.128742}
\showDOI{\tempurl}


\bibitem[Birman and Joseph(1987b)]%
        {birman-reliable}
\bibfield{author}{\bibinfo{person}{Kenneth~P. Birman} {and}
  \bibinfo{person}{Thomas~A. Joseph}.} \bibinfo{year}{1987}\natexlab{b}.
\newblock \showarticletitle{Reliable Communication in the Presence of
  Failures}.
\newblock \bibinfo{journal}{\emph{ACM Trans. Comput. Syst.}}
  \bibinfo{volume}{5}, \bibinfo{number}{1} (\bibinfo{date}{Jan.}
  \bibinfo{year}{1987}), \bibinfo{pages}{47–76}.
\newblock
\showISSN{0734-2071}
\urldef\tempurl%
\url{https://doi.org/10.1145/7351.7478}
\showDOI{\tempurl}


\bibitem[Bollig et~al\mbox{.}(2021)]%
        {bollig-mscs}
\bibfield{author}{\bibinfo{person}{Benedikt Bollig}, \bibinfo{person}{Cinzia
  Di~Giusto}, \bibinfo{person}{Alain Finkel}, \bibinfo{person}{Laetitia
  Laversa}, \bibinfo{person}{Etienne Lozes}, {and} \bibinfo{person}{Amrita
  Suresh}.} \bibinfo{year}{2021}\natexlab{}.
\newblock \showarticletitle{A Unifying Framework for Deciding
  Synchronizability}. In \bibinfo{booktitle}{\emph{32nd International
  Conference on Concurrency Theory (CONCUR 2021)}}
  \emph{(\bibinfo{series}{Leibniz International Proceedings in Informatics
  (LIPIcs)}, Vol.~\bibinfo{volume}{203})},
  \bibfield{editor}{\bibinfo{person}{Serge Haddad} {and}
  \bibinfo{person}{Daniele Varacca}} (Eds.). \bibinfo{publisher}{Schloss
  Dagstuhl -- Leibniz-Zentrum f{\"u}r Informatik}, \bibinfo{address}{Dagstuhl,
  Germany}, \bibinfo{pages}{14:1--14:18}.
\newblock
\showISBNx{978-3-95977-203-7}
\showISSN{1868-8969}
\urldef\tempurl%
\url{https://doi.org/10.4230/LIPIcs.CONCUR.2021.14}
\showDOI{\tempurl}


\bibitem[Bonchi et~al\mbox{.}(2023)]%
        {bonchi-tape-diagrams}
\bibfield{author}{\bibinfo{person}{Filippo Bonchi}, \bibinfo{person}{Alessandro
  Di~Giorgio}, {and} \bibinfo{person}{Alessio Santamaria}.}
  \bibinfo{year}{2023}\natexlab{}.
\newblock \showarticletitle{Deconstructing the Calculus of Relations with Tape
  Diagrams}.
\newblock \bibinfo{journal}{\emph{Proc. ACM Program. Lang.}}
  \bibinfo{volume}{7}, \bibinfo{number}{{POPL}}, Article
  \bibinfo{articleno}{64} (\bibinfo{date}{Jan.} \bibinfo{year}{2023}),
  \bibinfo{numpages}{31}~pages.
\newblock
\urldef\tempurl%
\url{https://doi.org/10.1145/3571257}
\showDOI{\tempurl}


\bibitem[Brookes and O'Hearn(2016)]%
        {brookes-ohearn-csl}
\bibfield{author}{\bibinfo{person}{Stephen Brookes} {and}
  \bibinfo{person}{Peter~W. O'Hearn}.} \bibinfo{year}{2016}\natexlab{}.
\newblock \showarticletitle{Concurrent Separation Logic}.
\newblock \bibinfo{journal}{\emph{ACM SIGLOG News}} \bibinfo{volume}{3},
  \bibinfo{number}{3} (\bibinfo{date}{Aug.} \bibinfo{year}{2016}),
  \bibinfo{pages}{47–65}.
\newblock
\urldef\tempurl%
\url{https://doi.org/10.1145/2984450.2984457}
\showDOI{\tempurl}


\bibitem[Broy(2005)]%
        {broy-msc-semantics}
\bibfield{author}{\bibinfo{person}{Manfred Broy}.}
  \bibinfo{year}{2005}\natexlab{}.
\newblock \showarticletitle{A semantic and methodological essence of message
  sequence charts}.
\newblock \bibinfo{journal}{\emph{Science of Computer Programming}}
  \bibinfo{volume}{54}, \bibinfo{number}{2} (\bibinfo{year}{2005}),
  \bibinfo{pages}{213--256}.
\newblock
\showISSN{0167-6423}
\urldef\tempurl%
\url{https://doi.org/10.1016/j.scico.2004.04.003}
\showDOI{\tempurl}


\bibitem[Castro and Liskov(1999)]%
        {castro-liskov-pbft}
\bibfield{author}{\bibinfo{person}{Miguel Castro} {and}
  \bibinfo{person}{Barbara Liskov}.} \bibinfo{year}{1999}\natexlab{}.
\newblock \showarticletitle{Practical Byzantine Fault Tolerance}. In
  \bibinfo{booktitle}{\emph{Proceedings of the Third Symposium on Operating
  Systems Design and Implementation}} (New Orleans, Louisiana, USA)
  \emph{(\bibinfo{series}{OSDI '99})}. \bibinfo{publisher}{USENIX Association},
  \bibinfo{address}{USA}, \bibinfo{pages}{173–186}.
\newblock
\showISBNx{1880446391}


\bibitem[Chandy and Lamport(1985)]%
        {chandy-lamport-snapshots}
\bibfield{author}{\bibinfo{person}{K.~Mani Chandy} {and}
  \bibinfo{person}{Leslie Lamport}.} \bibinfo{year}{1985}\natexlab{}.
\newblock \showarticletitle{Distributed Snapshots: Determining Global States of
  Distributed Systems}.
\newblock \bibinfo{journal}{\emph{ACM Trans. Comput. Syst.}}
  \bibinfo{volume}{3}, \bibinfo{number}{1} (\bibinfo{date}{Feb.}
  \bibinfo{year}{1985}), \bibinfo{pages}{63–75}.
\newblock
\showISSN{0734-2071}
\urldef\tempurl%
\url{https://doi.org/10.1145/214451.214456}
\showDOI{\tempurl}


\bibitem[Coecke and Duncan(2008)]%
        {coecke-zx-calculus}
\bibfield{author}{\bibinfo{person}{Bob Coecke} {and} \bibinfo{person}{Ross
  Duncan}.} \bibinfo{year}{2008}\natexlab{}.
\newblock \showarticletitle{Interacting Quantum Observables}. In
  \bibinfo{booktitle}{\emph{Automata, Languages and Programming}},
  \bibfield{editor}{\bibinfo{person}{Luca Aceto}, \bibinfo{person}{Ivan
  Damg{\aa}rd}, \bibinfo{person}{Leslie~Ann Goldberg},
  \bibinfo{person}{Magn{\'u}s~M. Halld{\'o}rsson}, \bibinfo{person}{Anna
  Ing{\'o}lfsd{\'o}ttir}, {and} \bibinfo{person}{Igor Walukiewicz}} (Eds.).
  \bibinfo{publisher}{Springer Berlin Heidelberg}, \bibinfo{address}{Berlin,
  Heidelberg}, \bibinfo{pages}{298--310}.
\newblock
\showISBNx{978-3-540-70583-3}


\bibitem[Di~Giusto et~al\mbox{.}(2023)]%
        {di-giusto-mscs}
\bibfield{author}{\bibinfo{person}{Cinzia Di~Giusto}, \bibinfo{person}{Davide
  Ferr\'{e}}, \bibinfo{person}{Laetitia Laversa}, {and}
  \bibinfo{person}{Etienne Lozes}.} \bibinfo{year}{2023}\natexlab{}.
\newblock \showarticletitle{A Partial Order View of Message-Passing
  Communication Models}.
\newblock \bibinfo{journal}{\emph{Proc. ACM Program. Lang.}}
  \bibinfo{volume}{7}, \bibinfo{number}{{POPL}}, Article
  \bibinfo{articleno}{55} (\bibinfo{date}{Jan.} \bibinfo{year}{2023}),
  \bibinfo{numpages}{27}~pages.
\newblock
\urldef\tempurl%
\url{https://doi.org/10.1145/3571248}
\showDOI{\tempurl}


\bibitem[Ellis and Gibbs(1989)]%
        {ellis-gibbs-ot}
\bibfield{author}{\bibinfo{person}{C.~A. Ellis} {and} \bibinfo{person}{S.~J.
  Gibbs}.} \bibinfo{year}{1989}\natexlab{}.
\newblock \showarticletitle{Concurrency Control in Groupware Systems}.
\newblock \bibinfo{journal}{\emph{SIGMOD Rec.}} \bibinfo{volume}{18},
  \bibinfo{number}{2} (\bibinfo{date}{June} \bibinfo{year}{1989}),
  \bibinfo{pages}{399–407}.
\newblock
\showISSN{0163-5808}
\urldef\tempurl%
\url{https://doi.org/10.1145/66926.66963}
\showDOI{\tempurl}


\bibitem[Fidge(1988)]%
        {fidge-vector-time}
\bibfield{author}{\bibinfo{person}{C.~J. Fidge}.}
  \bibinfo{year}{1988}\natexlab{}.
\newblock \showarticletitle{Timestamps in message-passing systems that preserve
  the partial ordering}.
\newblock \bibinfo{journal}{\emph{Proceedings of the 11th Australian Computer
  Science Conference}} \bibinfo{volume}{10}, \bibinfo{number}{1}
  (\bibinfo{year}{1988}), \bibinfo{pages}{56–66}.
\newblock


\bibitem[Flanagan and Freund(2009)]%
        {flanagan-fasttrack}
\bibfield{author}{\bibinfo{person}{Cormac Flanagan} {and}
  \bibinfo{person}{Stephen~N. Freund}.} \bibinfo{year}{2009}\natexlab{}.
\newblock \showarticletitle{FastTrack: Efficient and Precise Dynamic Race
  Detection}. In \bibinfo{booktitle}{\emph{Proceedings of the 30th ACM SIGPLAN
  Conference on Programming Language Design and Implementation}} (Dublin,
  Ireland) \emph{(\bibinfo{series}{PLDI '09})}. \bibinfo{publisher}{Association
  for Computing Machinery}, \bibinfo{address}{New York, NY, USA},
  \bibinfo{pages}{121–133}.
\newblock
\showISBNx{9781605583921}
\urldef\tempurl%
\url{https://doi.org/10.1145/1542476.1542490}
\showDOI{\tempurl}


\bibitem[Gehrke et~al\mbox{.}(1998)]%
        {gehrke-msc-semantics}
\bibfield{author}{\bibinfo{person}{Thomas Gehrke}, \bibinfo{person}{Michaela
  Huhn}, \bibinfo{person}{Arend Rensink}, {and} \bibinfo{person}{Heike
  Wehrheim}.} \bibinfo{year}{1998}\natexlab{}.
\newblock \bibinfo{booktitle}{\emph{An Algebraic Semantics for Message Sequence
  Chart Documents}}.
\newblock \bibinfo{publisher}{Springer US}, \bibinfo{address}{Boston, MA},
  \bibinfo{pages}{3--18}.
\newblock
\showISBNx{978-0-387-35394-4}
\urldef\tempurl%
\url{https://doi.org/10.1007/978-0-387-35394-4_1}
\showDOI{\tempurl}


\bibitem[Gondelman et~al\mbox{.}(2021)]%
        {gondelman-distributed-causal-memory}
\bibfield{author}{\bibinfo{person}{L\'{e}on Gondelman},
  \bibinfo{person}{Simon~Oddershede Gregersen}, \bibinfo{person}{Abel Nieto},
  \bibinfo{person}{Amin Timany}, {and} \bibinfo{person}{Lars Birkedal}.}
  \bibinfo{year}{2021}\natexlab{}.
\newblock \showarticletitle{Distributed Causal Memory: Modular Specification
  and Verification in Higher-Order Distributed Separation Logic}.
\newblock \bibinfo{journal}{\emph{Proc. ACM Program. Lang.}}
  \bibinfo{volume}{5}, \bibinfo{number}{{POPL}}, Article
  \bibinfo{articleno}{42} (\bibinfo{date}{Jan.} \bibinfo{year}{2021}),
  \bibinfo{numpages}{29}~pages.
\newblock
\urldef\tempurl%
\url{https://doi.org/10.1145/3434323}
\showDOI{\tempurl}


\bibitem[{ITU-T}(2011)]%
        {msc-standard}
\bibfield{author}{\bibinfo{person}{{ITU-T}}.} \bibinfo{year}{2011}\natexlab{}.
\newblock \bibinfo{title}{ITU Recommendation Z.120: Message Sequence Chart
  (MSC)}.
\newblock
\newblock
\urldef\tempurl%
\url{https://www.itu.int/rec/T-REC-Z.120-201102-I/}
\showURL{%
\tempurl}


\bibitem[Joyal and Street(1991)]%
        {joyal-street-geometry}
\bibfield{author}{\bibinfo{person}{André Joyal} {and} \bibinfo{person}{Ross
  Street}.} \bibinfo{year}{1991}\natexlab{}.
\newblock \showarticletitle{The geometry of tensor calculus, I}.
\newblock \bibinfo{journal}{\emph{Advances in Mathematics}}
  \bibinfo{volume}{88}, \bibinfo{number}{1} (\bibinfo{date}{July}
  \bibinfo{year}{1991}), \bibinfo{pages}{55--112}.
\newblock
\showISSN{0001-8708}
\urldef\tempurl%
\url{https://doi.org/10.1016/0001-8708(91)90003-p}
\showDOI{\tempurl}


\bibitem[Jung et~al\mbox{.}(2018)]%
        {jung-iris}
\bibfield{author}{\bibinfo{person}{Ralf Jung}, \bibinfo{person}{Robbert
  Krebbers}, \bibinfo{person}{Jacques{-}Henri Jourdan}, \bibinfo{person}{Ales
  Bizjak}, \bibinfo{person}{Lars Birkedal}, {and} \bibinfo{person}{Derek
  Dreyer}.} \bibinfo{year}{2018}\natexlab{}.
\newblock \showarticletitle{Iris from the ground up: {A} modular foundation for
  higher-order concurrent separation logic}.
\newblock \bibinfo{journal}{\emph{J. Funct. Program.}}  \bibinfo{volume}{28}
  (\bibinfo{year}{2018}), \bibinfo{pages}{e20}.
\newblock
\urldef\tempurl%
\url{https://doi.org/10.1017/S0956796818000151}
\showDOI{\tempurl}


\bibitem[Krogh-Jespersen et~al\mbox{.}(2020)]%
        {krogh-jespersen-aneris}
\bibfield{author}{\bibinfo{person}{Morten Krogh-Jespersen},
  \bibinfo{person}{Amin Timany}, \bibinfo{person}{Marit~Edna Ohlenbusch},
  \bibinfo{person}{Simon~Oddershede Gregersen}, {and} \bibinfo{person}{Lars
  Birkedal}.} \bibinfo{year}{2020}\natexlab{}.
\newblock \showarticletitle{Aneris: A Mechanised Logic for Modular Reasoning
  about Distributed Systems}. In \bibinfo{booktitle}{\emph{Programming
  Languages and Systems: 29th European Symposium on Programming, ESOP 2020,
  Held as Part of the European Joint Conferences on Theory and Practice of
  Software, ETAPS 2020, Dublin, Ireland, April 25–30, 2020, Proceedings}}
  (Dublin, Ireland). \bibinfo{publisher}{Springer-Verlag},
  \bibinfo{address}{Berlin, Heidelberg}, \bibinfo{pages}{336–365}.
\newblock
\showISBNx{978-3-030-44913-1}
\urldef\tempurl%
\url{https://doi.org/10.1007/978-3-030-44914-8_13}
\showDOI{\tempurl}


\bibitem[Kshemkalyani(2004)]%
        {kshemkalyani-tensor-clocks}
\bibfield{author}{\bibinfo{person}{Ajay~D. Kshemkalyani}.}
  \bibinfo{year}{2004}\natexlab{}.
\newblock \showarticletitle{The power of logical clock abstractions}.
\newblock \bibinfo{journal}{\emph{Distributed Computing}} \bibinfo{volume}{17},
  \bibinfo{number}{2} (\bibinfo{date}{Aug.} \bibinfo{year}{2004}).
\newblock
\showISSN{0178-2770, 1432-0452}
\urldef\tempurl%
\url{https://doi.org/10.1007/s00446-003-0105-9}
\showDOI{\tempurl}


\bibitem[Ladkin and Leue(1993)]%
        {ladkin-msc-semantics}
\bibfield{author}{\bibinfo{person}{Peter~B. Ladkin} {and}
  \bibinfo{person}{Stefan Leue}.} \bibinfo{year}{1993}\natexlab{}.
\newblock \showarticletitle{What Do Message Sequence Charts Mean?}. In
  \bibinfo{booktitle}{\emph{Proceedings of the IFIP TC6/WG6.1 Sixth
  International Conference on Formal Description Techniques, VI}}
  \emph{(\bibinfo{series}{FORTE '93})}. \bibinfo{publisher}{North-Holland
  Publishing Co.}, \bibinfo{address}{Nld}, \bibinfo{pages}{301–316}.
\newblock
\showISBNx{0444817735}


\bibitem[Lamport(1978)]%
        {lamport-clocks}
\bibfield{author}{\bibinfo{person}{Leslie Lamport}.}
  \bibinfo{year}{1978}\natexlab{}.
\newblock \showarticletitle{Time, Clocks, and the Ordering of Events in a
  Distributed System}.
\newblock \bibinfo{journal}{\emph{Commun. ACM}} \bibinfo{volume}{21},
  \bibinfo{number}{7} (\bibinfo{date}{July} \bibinfo{year}{1978}),
  \bibinfo{pages}{558--565}.
\newblock
\showISSN{0001-0782}
\urldef\tempurl%
\url{https://doi.org/10.1145/359545.359563}
\showDOI{\tempurl}


\bibitem[Le~Lann(1977)]%
        {le-lann-distsys}
\bibfield{author}{\bibinfo{person}{G\'erard Le~Lann}.}
  \bibinfo{year}{1977}\natexlab{}.
\newblock \showarticletitle{Distributed Systems -- Toward a Formal Approach}.
  In \bibinfo{booktitle}{\emph{Proceedings of IFIP Congress 1977}} (Toronto,
  Canada) \emph{(\bibinfo{series}{IFIP '77})}.
  \bibinfo{publisher}{North-Holland Publishing Co.}, \bibinfo{address}{Nld},
  \bibinfo{pages}{155--160}.
\newblock
\showISBNx{0720407559}


\bibitem[Lehmann et~al\mbox{.}(2022)]%
        {lehmann-vyzx}
\bibfield{author}{\bibinfo{person}{Adrian Lehmann}, \bibinfo{person}{Ben
  Caldwell}, {and} \bibinfo{person}{Robert Rand}.}
  \bibinfo{year}{2022}\natexlab{}.
\newblock \bibinfo{title}{{VyZX}: A Vision for Verifying the {ZX} Calculus}.
\newblock
\newblock
\urldef\tempurl%
\url{https://doi.org/10.48550/ARXIV.2205.05781}
\showDOI{\tempurl}
\showeprint[arxiv]{2205.05781}~[quant-ph]


\bibitem[Lehmann et~al\mbox{.}(2023)]%
        {lehmann-vyzx-2023}
\bibfield{author}{\bibinfo{person}{Adrian Lehmann}, \bibinfo{person}{Ben
  Caldwell}, \bibinfo{person}{Bhakti Shah}, {and} \bibinfo{person}{Robert
  Rand}.} \bibinfo{year}{2023}\natexlab{}.
\newblock \bibinfo{title}{VyZX: Formal Verification of a Graphical Quantum
  Language}.
\newblock
\newblock
\urldef\tempurl%
\url{https://doi.org/10.48550/arXiv.2311.11571}
\showDOI{\tempurl}
\showeprint[arxiv]{2311.11571}~[cs.PL]


\bibitem[Lesani et~al\mbox{.}(2016)]%
        {lesani-chapar}
\bibfield{author}{\bibinfo{person}{Mohsen Lesani},
  \bibinfo{person}{Christian~J. Bell}, {and} \bibinfo{person}{Adam Chlipala}.}
  \bibinfo{year}{2016}\natexlab{}.
\newblock \showarticletitle{Chapar: Certified Causally Consistent Distributed
  Key-Value Stores}. In \bibinfo{booktitle}{\emph{Proceedings of the 43rd
  Annual ACM SIGPLAN-SIGACT Symposium on Principles of Programming Languages}}
  (St. Petersburg, FL, USA) \emph{(\bibinfo{series}{{POPL '16}})}.
  \bibinfo{publisher}{Association for Computing Machinery},
  \bibinfo{address}{New York, NY, USA}, \bibinfo{pages}{357–370}.
\newblock
\showISBNx{9781450335492}
\urldef\tempurl%
\url{https://doi.org/10.1145/2837614.2837622}
\showDOI{\tempurl}


\bibitem[Lloyd et~al\mbox{.}(2011)]%
        {lloyd-cops}
\bibfield{author}{\bibinfo{person}{Wyatt Lloyd}, \bibinfo{person}{Michael~J.
  Freedman}, \bibinfo{person}{Michael Kaminsky}, {and}
  \bibinfo{person}{David~G. Andersen}.} \bibinfo{year}{2011}\natexlab{}.
\newblock \showarticletitle{Don't Settle for Eventual: Scalable Causal
  Consistency for Wide-Area Storage with COPS}. In
  \bibinfo{booktitle}{\emph{Proceedings of the Twenty-Third ACM Symposium on
  Operating Systems Principles}} (Cascais, Portugal)
  \emph{(\bibinfo{series}{SOSP '11})}. \bibinfo{publisher}{Association for
  Computing Machinery}, \bibinfo{address}{New York, NY, USA},
  \bibinfo{pages}{401–416}.
\newblock
\showISBNx{9781450309776}
\urldef\tempurl%
\url{https://doi.org/10.1145/2043556.2043593}
\showDOI{\tempurl}


\bibitem[Lohrey and Muscholl(2004)]%
        {lohrey-mscs}
\bibfield{author}{\bibinfo{person}{Markus Lohrey} {and} \bibinfo{person}{Anca
  Muscholl}.} \bibinfo{year}{2004}\natexlab{}.
\newblock \showarticletitle{Bounded MSC communication}.
\newblock \bibinfo{journal}{\emph{Information and Computation}}
  \bibinfo{volume}{189}, \bibinfo{number}{2} (\bibinfo{year}{2004}),
  \bibinfo{pages}{160--181}.
\newblock
\showISSN{0890-5401}
\urldef\tempurl%
\url{https://doi.org/10.1016/j.ic.2003.10.002}
\showDOI{\tempurl}


\bibitem[Mansky et~al\mbox{.}(2017)]%
        {mansky-verified-clocks}
\bibfield{author}{\bibinfo{person}{William Mansky}, \bibinfo{person}{Yuanfeng
  Peng}, \bibinfo{person}{Steve Zdancewic}, {and} \bibinfo{person}{Joseph
  Devietti}.} \bibinfo{year}{2017}\natexlab{}.
\newblock \showarticletitle{Verifying Dynamic Race Detection}. In
  \bibinfo{booktitle}{\emph{Proceedings of the 6th ACM SIGPLAN Conference on
  Certified Programs and Proofs}} (Paris, France) \emph{(\bibinfo{series}{Cpp
  2017})}. \bibinfo{publisher}{Association for Computing Machinery},
  \bibinfo{address}{New York, NY, USA}, \bibinfo{pages}{151–163}.
\newblock
\showISBNx{9781450347051}
\urldef\tempurl%
\url{https://doi.org/10.1145/3018610.3018611}
\showDOI{\tempurl}


\bibitem[Mathur et~al\mbox{.}(2022)]%
        {mathur-tree-clocks}
\bibfield{author}{\bibinfo{person}{Umang Mathur}, \bibinfo{person}{Andreas
  Pavlogiannis}, \bibinfo{person}{H\"{u}nkar~Can Tun\c{c}}, {and}
  \bibinfo{person}{Mahesh Viswanathan}.} \bibinfo{year}{2022}\natexlab{}.
\newblock \showarticletitle{A Tree Clock Data Structure for Causal Orderings in
  Concurrent Executions}. In \bibinfo{booktitle}{\emph{Proceedings of the 27th
  ACM International Conference on Architectural Support for Programming
  Languages and Operating Systems}} (Lausanne, Switzerland)
  \emph{(\bibinfo{series}{ASPLOS '22})}. \bibinfo{publisher}{Association for
  Computing Machinery}, \bibinfo{address}{New York, NY, USA},
  \bibinfo{pages}{710–725}.
\newblock
\showISBNx{9781450392051}
\urldef\tempurl%
\url{https://doi.org/10.1145/3503222.3507734}
\showDOI{\tempurl}


\bibitem[Mattern(1989)]%
        {mattern-vector-time}
\bibfield{author}{\bibinfo{person}{Friedemann Mattern}.}
  \bibinfo{year}{1989}\natexlab{}.
\newblock \showarticletitle{Virtual Time and Global States of Distributed
  Systems}. In \bibinfo{booktitle}{\emph{Parallel and Distributed Algorithms}}.
  \bibinfo{publisher}{North-Holland}, \bibinfo{pages}{215--226}.
\newblock


\bibitem[Mauw and Reniers(1994)]%
        {mauw-reniers-msc-semantics}
\bibfield{author}{\bibinfo{person}{S. Mauw} {and} \bibinfo{person}{M.~A.
  Reniers}.} \bibinfo{year}{1994}\natexlab{}.
\newblock \showarticletitle{An Algebraic Semantics of Basic Message Sequence
  Charts}.
\newblock \bibinfo{journal}{\emph{Comput. J.}} \bibinfo{volume}{37},
  \bibinfo{number}{4} (\bibinfo{date}{01} \bibinfo{year}{1994}),
  \bibinfo{pages}{269--277}.
\newblock
\showISSN{0010-4620}
\urldef\tempurl%
\url{https://doi.org/10.1093/comjnl/37.4.269}
\showDOI{\tempurl}


\bibitem[Mora et~al\mbox{.}(2023)]%
        {mora-message-chains}
\bibfield{author}{\bibinfo{person}{Federico Mora}, \bibinfo{person}{Ankush
  Desai}, \bibinfo{person}{Elizabeth Polgreen}, {and}
  \bibinfo{person}{Sanjit~A. Seshia}.} \bibinfo{year}{2023}\natexlab{}.
\newblock \showarticletitle{Message Chains for Distributed System
  Verification}.
\newblock \bibinfo{journal}{\emph{Proc. ACM Program. Lang.}}
  \bibinfo{volume}{7}, \bibinfo{number}{OOPSLA2}, Article
  \bibinfo{articleno}{300} (\bibinfo{date}{Oct.} \bibinfo{year}{2023}),
  \bibinfo{numpages}{27}~pages.
\newblock
\urldef\tempurl%
\url{https://doi.org/10.1145/3622876}
\showDOI{\tempurl}


\bibitem[Nester(2021)]%
        {nester-concurrent-histories}
\bibfield{author}{\bibinfo{person}{Chad Nester}.}
  \bibinfo{year}{2021}\natexlab{}.
\newblock \showarticletitle{The Structure of Concurrent Process Histories}. In
  \bibinfo{booktitle}{\emph{Coordination {Models} and {Languages}}}
  \emph{(\bibinfo{series}{Lecture {Notes} in {Computer} {Science}})},
  \bibfield{editor}{\bibinfo{person}{Ferruccio Damiani} {and}
  \bibinfo{person}{Ornela Dardha}} (Eds.). \bibinfo{publisher}{Springer
  International Publishing}, \bibinfo{address}{Cham},
  \bibinfo{pages}{209--224}.
\newblock
\showISBNx{9783030781422}
\urldef\tempurl%
\url{https://doi.org/10.1007/978-3-030-78142-2_13}
\showDOI{\tempurl}


\bibitem[Nieto et~al\mbox{.}(2022)]%
        {nieto-verified-crdts}
\bibfield{author}{\bibinfo{person}{Abel Nieto}, \bibinfo{person}{L\'{e}on
  Gondelman}, \bibinfo{person}{Alban Reynaud}, \bibinfo{person}{Amin Timany},
  {and} \bibinfo{person}{Lars Birkedal}.} \bibinfo{year}{2022}\natexlab{}.
\newblock \showarticletitle{Modular Verification of Op-Based CRDTs in
  Separation Logic}.
\newblock \bibinfo{journal}{\emph{Proc. ACM Program. Lang.}}
  \bibinfo{volume}{6}, \bibinfo{number}{OOPSLA2}, Article
  \bibinfo{articleno}{188} (\bibinfo{date}{Oct.} \bibinfo{year}{2022}),
  \bibinfo{numpages}{29}~pages.
\newblock
\urldef\tempurl%
\url{https://doi.org/10.1145/3563351}
\showDOI{\tempurl}


\bibitem[O'Hearn(2019)]%
        {ohearn-cacm-sl}
\bibfield{author}{\bibinfo{person}{Peter O'Hearn}.}
  \bibinfo{year}{2019}\natexlab{}.
\newblock \showarticletitle{Separation Logic}.
\newblock \bibinfo{journal}{\emph{Commun. ACM}} \bibinfo{volume}{62},
  \bibinfo{number}{2} (\bibinfo{date}{Jan.} \bibinfo{year}{2019}),
  \bibinfo{pages}{86–95}.
\newblock
\showISSN{0001-0782}
\urldef\tempurl%
\url{https://doi.org/10.1145/3211968}
\showDOI{\tempurl}


\bibitem[O’Hearn(2007)]%
        {ohearn-csl}
\bibfield{author}{\bibinfo{person}{Peter~W. O’Hearn}.}
  \bibinfo{year}{2007}\natexlab{}.
\newblock \showarticletitle{Resources, concurrency, and local reasoning}.
\newblock \bibinfo{journal}{\emph{Theoretical Computer Science}}
  \bibinfo{volume}{375}, \bibinfo{number}{1} (\bibinfo{year}{2007}),
  \bibinfo{pages}{271--307}.
\newblock
\showISSN{0304-3975}
\urldef\tempurl%
\url{https://doi.org/10.1016/j.tcs.2006.12.035}
\showDOI{\tempurl}
\newblock
\shownote{Festschrift for John C. Reynolds’s 70th birthday}.


\bibitem[Piedeleu and Zanasi(2023)]%
        {piedeleu-intro-string-diagrams}
\bibfield{author}{\bibinfo{person}{Robin Piedeleu} {and} \bibinfo{person}{Fabio
  Zanasi}.} \bibinfo{year}{2023}\natexlab{}.
\newblock \bibinfo{title}{An Introduction to String Diagrams for Computer
  Scientists}.
\newblock
\newblock
\showeprint[arxiv]{2305.08768}~[cs.LO]
\urldef\tempurl%
\url{https://arxiv.org/abs/2305.08768}
\showURL{%
\tempurl}


\bibitem[Pozniansky and Schuster(2003)]%
        {pozniansky-djit}
\bibfield{author}{\bibinfo{person}{Eli Pozniansky} {and} \bibinfo{person}{Assaf
  Schuster}.} \bibinfo{year}{2003}\natexlab{}.
\newblock \showarticletitle{Efficient On-the-Fly Data Race Detection in
  Multithreaded C++ Programs}. In \bibinfo{booktitle}{\emph{Proceedings of the
  Ninth ACM SIGPLAN Symposium on Principles and Practice of Parallel
  Programming}} (San Diego, California, USA) \emph{(\bibinfo{series}{PPoPP
  '03})}. \bibinfo{publisher}{Association for Computing Machinery},
  \bibinfo{address}{New York, NY, USA}, \bibinfo{pages}{179–190}.
\newblock
\showISBNx{1581135882}
\urldef\tempurl%
\url{https://doi.org/10.1145/781498.781529}
\showDOI{\tempurl}


\bibitem[Rahli et~al\mbox{.}(2017)]%
        {rahli-eventml}
\bibfield{author}{\bibinfo{person}{Vincent Rahli}, \bibinfo{person}{David
  Guaspari}, \bibinfo{person}{Mark Bickford}, {and} \bibinfo{person}{Robert~L.
  Constable}.} \bibinfo{year}{2017}\natexlab{}.
\newblock \showarticletitle{EventML: Specification, verification, and
  implementation of crash-tolerant state machine replication systems}.
\newblock \bibinfo{journal}{\emph{Science of Computer Programming}}
  \bibinfo{volume}{148} (\bibinfo{year}{2017}), \bibinfo{pages}{26--48}.
\newblock
\showISSN{0167-6423}
\urldef\tempurl%
\url{https://doi.org/10.1016/j.scico.2017.05.009}
\showDOI{\tempurl}
\newblock
\shownote{Special issue on Automated Verification of Critical Systems (AVoCS
  2015)}.


\bibitem[Raynal et~al\mbox{.}(1991)]%
        {rst-matrix-clock}
\bibfield{author}{\bibinfo{person}{Michel Raynal}, \bibinfo{person}{André
  Schiper}, {and} \bibinfo{person}{Sam Toueg}.}
  \bibinfo{year}{1991}\natexlab{}.
\newblock \showarticletitle{The causal ordering abstraction and a simple way to
  implement it}.
\newblock \bibinfo{journal}{\emph{Inform. Process. Lett.}}
  \bibinfo{volume}{39}, \bibinfo{number}{6} (\bibinfo{date}{Sept.}
  \bibinfo{year}{1991}), \bibinfo{pages}{343--350}.
\newblock
\showISSN{00200190}
\urldef\tempurl%
\url{https://doi.org/10.1016/0020-0190(91)90008-6}
\showDOI{\tempurl}


\bibitem[Raynal and Singhal(1996)]%
        {raynal-singhal-survey}
\bibfield{author}{\bibinfo{person}{Michel Raynal} {and} \bibinfo{person}{Mukesh
  Singhal}.} \bibinfo{year}{1996}\natexlab{}.
\newblock \showarticletitle{Logical time: Capturing causality in distributed
  systems}.
\newblock \bibinfo{journal}{\emph{Computer}} \bibinfo{volume}{29},
  \bibinfo{number}{2} (\bibinfo{year}{1996}), \bibinfo{pages}{49--56}.
\newblock


\bibitem[Redmond et~al\mbox{.}(2023)]%
        {redmond-verified-causal-broadcast}
\bibfield{author}{\bibinfo{person}{Patrick Redmond}, \bibinfo{person}{Gan
  Shen}, \bibinfo{person}{Niki Vazou}, {and} \bibinfo{person}{Lindsey Kuper}.}
  \bibinfo{year}{2023}\natexlab{}.
\newblock \showarticletitle{Verified Causal Broadcast with Liquid Haskell}. In
  \bibinfo{booktitle}{\emph{Proceedings of the 34th Symposium on Implementation
  and Application of Functional Languages}} (Copenhagen, Denmark)
  \emph{(\bibinfo{series}{IFL '22})}. \bibinfo{publisher}{Association for
  Computing Machinery}, \bibinfo{address}{New York, NY, USA}, Article
  \bibinfo{articleno}{6}, \bibinfo{numpages}{13}~pages.
\newblock
\showISBNx{9781450398312}
\urldef\tempurl%
\url{https://doi.org/10.1145/3587216.3587222}
\showDOI{\tempurl}


\bibitem[Reynolds(2002)]%
        {reynolds-separation-logic}
\bibfield{author}{\bibinfo{person}{J.C. Reynolds}.}
  \bibinfo{year}{2002}\natexlab{}.
\newblock \showarticletitle{Separation logic: a logic for shared mutable data
  structures}. In \bibinfo{booktitle}{\emph{Proceedings 17th Annual IEEE
  Symposium on Logic in Computer Science}}. \bibinfo{pages}{55--74}.
\newblock
\urldef\tempurl%
\url{https://doi.org/10.1109/LICS.2002.1029817}
\showDOI{\tempurl}


\bibitem[Sch{\"a}tz et~al\mbox{.}(1996)]%
        {schaltz-eets}
\bibfield{author}{\bibinfo{person}{Bernhard Sch{\"a}tz},
  \bibinfo{person}{Heinrich Hu{\ss}mann}, {and} \bibinfo{person}{Manfred
  Broy}.} \bibinfo{year}{1996}\natexlab{}.
\newblock \showarticletitle{Graphical development of consistent system
  specifications}. In \bibinfo{booktitle}{\emph{FME'96: Industrial Benefit and
  Advances in Formal Methods}}, \bibfield{editor}{\bibinfo{person}{Marie-Claude
  Gaudel} {and} \bibinfo{person}{James Woodcock}} (Eds.).
  \bibinfo{publisher}{Springer Berlin Heidelberg}, \bibinfo{address}{Berlin,
  Heidelberg}, \bibinfo{pages}{248--267}.
\newblock
\showISBNx{978-3-540-49749-3}


\bibitem[Schiper et~al\mbox{.}(1989)]%
        {schiper-causal-ordering}
\bibfield{author}{\bibinfo{person}{Andr\'{e} Schiper}, \bibinfo{person}{Jorge
  Eggli}, {and} \bibinfo{person}{Alain Sandoz}.}
  \bibinfo{year}{1989}\natexlab{}.
\newblock \showarticletitle{A New Algorithm to Implement Causal Ordering}. In
  \bibinfo{booktitle}{\emph{Proceedings of the 3rd International Workshop on
  Distributed Algorithms}}. \bibinfo{publisher}{Springer-Verlag},
  \bibinfo{address}{Berlin, Heidelberg}, \bibinfo{pages}{219–232}.
\newblock
\showISBNx{3540516875}


\bibitem[Schmuck(1988)]%
        {schmuck-dissertation}
\bibfield{author}{\bibinfo{person}{Frank~B Schmuck}.}
  \bibinfo{year}{1988}\natexlab{}.
\newblock \emph{\bibinfo{title}{The use of efficient broadcast protocols in
  asynchronous distributed systems}}.
\newblock \bibinfo{thesistype}{Ph.\,D. Dissertation}.
\newblock


\bibitem[Shapiro et~al\mbox{.}(2011)]%
        {shapiro-crdts}
\bibfield{author}{\bibinfo{person}{Marc Shapiro}, \bibinfo{person}{Nuno
  Pregui\c{c}a}, \bibinfo{person}{Carlos Baquero}, {and} \bibinfo{person}{Marek
  Zawirski}.} \bibinfo{year}{2011}\natexlab{}.
\newblock \showarticletitle{Conflict-Free Replicated Data Types}. In
  \bibinfo{booktitle}{\emph{Proceedings of the 13th International Conference on
  Stabilization, Safety, and Security of Distributed Systems}} (Grenoble,
  France) \emph{(\bibinfo{series}{SSS'11})}.
  \bibinfo{publisher}{Springer-Verlag}, \bibinfo{address}{Berlin, Heidelberg},
  \bibinfo{pages}{386–400}.
\newblock
\showISBNx{9783642245497}


\bibitem[Talupur and Tuttle(2008)]%
        {talupur-tuttle-message-flows}
\bibfield{author}{\bibinfo{person}{Murali Talupur} {and}
  \bibinfo{person}{Mark~R. Tuttle}.} \bibinfo{year}{2008}\natexlab{}.
\newblock \showarticletitle{Going with the Flow: Parameterized Verification
  Using Message Flows}. In \bibinfo{booktitle}{\emph{2008 Formal Methods in
  Computer-Aided Design}}. \bibinfo{pages}{1--8}.
\newblock
\urldef\tempurl%
\url{https://doi.org/10.1109/fmcad.2008.ecp.14}
\showDOI{\tempurl}


\bibitem[Weil et~al\mbox{.}(2006)]%
        {weil-ceph}
\bibfield{author}{\bibinfo{person}{Sage~A. Weil}, \bibinfo{person}{Scott~A.
  Brandt}, \bibinfo{person}{Ethan~L. Miller}, \bibinfo{person}{Darrell D.~E.
  Long}, {and} \bibinfo{person}{Carlos Maltzahn}.}
  \bibinfo{year}{2006}\natexlab{}.
\newblock \showarticletitle{Ceph: A Scalable, High-Performance Distributed File
  System}. In \bibinfo{booktitle}{\emph{Proceedings of the 7th Symposium on
  Operating Systems Design and Implementation}} (Seattle, Washington)
  \emph{(\bibinfo{series}{OSDI '06})}. \bibinfo{publisher}{USENIX Association},
  \bibinfo{address}{USA}, \bibinfo{pages}{307–320}.
\newblock
\showISBNx{1931971471}


\bibitem[Wickerson et~al\mbox{.}(2013)]%
        {wickerson-ribbon-proofs}
\bibfield{author}{\bibinfo{person}{John Wickerson}, \bibinfo{person}{Mike
  Dodds}, {and} \bibinfo{person}{Matthew~J. Parkinson}.}
  \bibinfo{year}{2013}\natexlab{}.
\newblock \showarticletitle{Ribbon Proofs for Separation Logic}. In
  \bibinfo{booktitle}{\emph{Programming Languages and Systems - 22nd European
  Symposium on Programming, {ESOP} 2013, Held as Part of the European Joint
  Conferences on Theory and Practice of Software, {ETAPS} 2013, Rome, Italy,
  March 16-24, 2013. Proceedings}} \emph{(\bibinfo{series}{Lecture Notes in
  Computer Science}, Vol.~\bibinfo{volume}{7792})},
  \bibfield{editor}{\bibinfo{person}{Matthias Felleisen} {and}
  \bibinfo{person}{Philippa Gardner}} (Eds.). \bibinfo{publisher}{Springer},
  \bibinfo{pages}{189--208}.
\newblock
\urldef\tempurl%
\url{https://doi.org/10.1007/978-3-642-37036-6\_12}
\showDOI{\tempurl}


\bibitem[Wuu and Bernstein(1984)]%
        {wuu-bernstein-matrix-clocks}
\bibfield{author}{\bibinfo{person}{Gene~T.J. Wuu} {and}
  \bibinfo{person}{Arthur~J. Bernstein}.} \bibinfo{year}{1984}\natexlab{}.
\newblock \showarticletitle{Efficient Solutions to the Replicated Log and
  Dictionary Problems}. In \bibinfo{booktitle}{\emph{Proceedings of the Third
  Annual ACM Symposium on Principles of Distributed Computing}} (Vancouver,
  British Columbia, Canada) \emph{(\bibinfo{series}{PODC '84})}.
  \bibinfo{publisher}{Association for Computing Machinery},
  \bibinfo{address}{New York, NY, USA}, \bibinfo{pages}{233–242}.
\newblock
\showISBNx{0897911431}
\urldef\tempurl%
\url{https://doi.org/10.1145/800222.806750}
\showDOI{\tempurl}


\end{thebibliography}

\end{document}